# Clifford Algebra-Based Iterated Extended Kalman Filter with Application to Low-Cost INS/GNSS Navigation

Wei Ouyang, Member IEEE, Yutian Wang, Yuanxin Wu, Senior Member, IEEE

*Abstract*— The traditional GNSS-aided inertial navigation system (INS) usually exploits the extended Kalman filter (EKF) for state estimation, and the initial attitude accuracy is key to the filtering performance. To spare the reliance on the initial attitude, this work generalizes the previously proposed trident quaternion within the framework of Clifford algebra to represent the extended pose, IMU biases and lever arms on the Lie group. Consequently, a quasi-group-affine system is established for the low-cost INS/GNSS integrated navigation system, and the right-error Clifford algebra-based EKF (Clifford-RQEKF) is accordingly developed. The iterated filtering approach is further applied to significantly improve the performances of the Clifford-RQEKF and the previously proposed trident quaternion-based EKFs. Numerical simulations and experiments show that all iterated filtering approaches fulfill the fast and global convergence without the prior attitude information, whereas the iterated Clifford-RQEKF performs much better than the others under especially large IMU biases.

## I. INTRODUCTION

Nowadays the Global Navigation Satellite System (GNSS) is nearly ubiquitous to users except for occasional signal blockages or interruptions by the surrounding buildings and trees in urban areas [1]-[3]. In these cases, the self-contained inertial navigation system (INS) can work in the dead-reckoning mode, continuing to provide complementary navigation information during the loss of the GNSS signal [4]. The low-cost INS/GNSS integrated navigation systems have been widely applied in the autonomous navigation of the land and aerial vehicles [5]-[9]. The navigation accuracy and efficiency of the INS/GNSS system mainly hinge on the underpinning state estimation algorithm.

The INS/GNSS integrated navigation systems typically use the extended Kalman filter (EKF) as the workhorse owning to its high efficiency and simplicity [1], [10], but the performance of EKF primarily depends on the accuracy of the initial attitude and observations. Recent years have witnessed the booming studies of optimization-based methods for INS/GNSS integrated navigation, which are largely implemented by the sliding-window optimization [11], [12] or the popular factor-graph optimization toolbox [13]-[17]. In contrast with the filtering-based methods, the estimation accuracy and robustness of optimization-based methods are superior, especially when the state is poorly initialized and there exist significant measurement outliers [10], [18]. Nevertheless, the complexity and computational costs of optimization-based methods are unamiable to the low-cost INS/GNSS integrated navigation system.

In the last decade, researchers have been endeavoring to enlarge the convergence domain of the Kalman filtering methods involving the attitude estimation. The main spirit is to minimize the linearization effect on the error evolution of the system kinematics. The common-frame filter GEKF [19] adopted the coordinate-consistent error representation for the gyroscope bias error, exhibiting improved accuracy and consistency in spacecraft attitude estimation under large initial attitude errors, and similar thoughts were applied to INS [20]-[22]. Moreover, the geometric properties of the rigid-body motion on the Special Euclidean group were also investigated for the inertial navigation problem. The striking invariant filtering theory proposed by Barrau and Bonnabel [23], [24] established the connection of system group-affine (trajectory-independent) condition with the estimation consistency and observability [25], [26], endowing the invariant EKF (IEKF) with the ability of local asymptotical convergence under large initial attitude errors. Related works such as Chang [27], [28], Cui [29] and Luo [30] showed that the invariant filtering methods based on the group-affine system perform much better than the standard EKF in INS/GNSS integrated navigation systems. From a more general perspective, the equivariant filter (EqF) recently proposed by van Goor [31] only requires the equivariance of the kinematic system posed on the homogeneous space, also minimizing the linearization error and showing superior performance in the biased attitude estimation [32] and visual-inertial odometry [33].

This paper was supported by in part by National Key R&D Program (2022YFB3903802), National Natural Science Foundation (62303310, 62273228). (Corresponding author: Yuanxin Wu).

Authors' address: The authors are with the institute of sensing and navigation, School of Electronic Information and Electrical Engineering, Shanghai Jiao Tong University, Shanghai 200240, China (email: ywoulife@sjtu.edu.cn, WangYutian_023@sjtu.edu.cn, yuanx_wu@hotmail.com).

### TABLE I
### Initial attitude errors considered in related works

|      | Bias (gyro, acc)         | Attitude error    |
|------|--------------------------|-------------------|
| [27] | 0.3 deg/h, 20 mg         | [60, 160, 60] deg |
| [28] | 0.3 deg/h, 20 mg         | [10, 179, 10] deg |
| [32] | ~0 deg/h [1], —          | 30 deg (std)      |
| [34] | ~200 deg/h [2], 0.04 mg  | 30 deg (std)      |
| [35] | 3600 deg/h, 0.1 g        | 30 deg (std)      |

[1]Calibrated gyro bias with in-run bias stability 1.8 deg/h; [2]In-run bias stability 5~7 deg/h (VN-100 high-performance IMU).

For the low-cost INS/GNSS system, however, the problem of significant IMU biases becomes a barrier to the group-affine condition, making the state-of-the-art methods "imperfect" in practice [34]. As indicated in [35], the system kinematic model is still dependent on the biases even when the system was modeled with the two-frames group (TFG) law. Moreover, the attitude estimation on SO(3) was unfeasible for the TFG-based IEKF in INS/GNSS integrated navigation because of the non-abelian property of SO(3) (see Section V.C of [35] for details). Due to significant IMU biases, the convergence region of the initial attitude error reduces to about 30 deg in standard deviations (std) for the state-of-the-art methods [32], [34], [35]. As compared in Table I, the invariant and equivariant filtering methods still cannot achieve the global convergence in the low-cost navigation system under arbitrarily initial attitude errors. Therefore, additional coarse alignment methods are still indispensable to acquire the initial attitude, generally obtained by the optimization-based [36], [37], Kalman filtering-based [38], [39] and trajectory matching-based methods [40], [41]. This flaw has baffled the practitioners in that the traditional EKF also works well with given fine initial attitude.

Apart from the adverse effect of IMU biases on the system model, the invariant linearization of the measurement model is also hard to be attained in practice. In [23]-[30], the lever arms between IMU and GNSS receiver/odometer are assumed to be calibrated to obtain the invariant linearization of the measurement model. Moreover, when the earth rotation is considered, the linearized measurement model is unavoidably state-dependent as shown in [27]-[30], [42].

In this article, a state estimation method for the low-cost INS/GNSS system free from the prior attitude is investigated. The recent work [43] indicated that the trident quaternion proposed in the companion work [44] is a subgroup of the Clifford algebra [45]-[47], and it is also isomorphic to the Lie group $SE_2(3)$ as shown herein. The extended pose along with the IMU biases and lever arms are expressed by a newly-designed Clifford algebra with explicit physical meanings. A quasi-group-affine kinematic model is formulated, and the corresponding right-error Clifford algebra-based EKF is derived. As a recursive form of the Gauss-Newton optimization, the iterated filtering technique [48], [49] is also used to minimize the residual linearization error in the measurement model. The main contributions are:

(1) The trident quaternion is defined as a subgroup of a Clifford algebra and proved to be isomorphic to the Lie group $SE_2(3)$. The Clifford algebra isomorphic to the Special Euclidean group $SE_{2+3}(3)$ is applied to represent the extended pose, IMU biases and lever arms.

(2) The quasi-group-affine system is constructed with the Clifford algebra using the right error, and the performances of the Clifford algebra-based EKFs are improved with iterated updates.

(3) Simulations and field experiments show that the proposed iterated Clifford-algebra EKFs converge quickly without requiring the initial attitude in the low-cost INS/GNSS integrated navigation system.

The remaining contents are organized as follows: Section II introduces the preliminaries of Clifford algebra. Section III drives the quasi-group-affine system using the Clifford algebra, and the corresponding right-error model is developed. Section IV compares the characteristics of the proposed Clifford-RQEKF with other invariant filtering methods, and the iterated filtering technique is introduced. The numerical simulations and experimental data tests are conducted in Section V. Section VI concludes this article.

## II. PRELIMINARIES OF THE CLIFFORD ALGEBRA

The fundamentals of Clifford algebra involving with the rigid-body motion representation has been given in the previous work [43] according to classical textbooks [45]-[47]. This section introduces the geometric product and the definition of the concerned Clifford algebra in detail.

### A. Clifford Algebra and Geometric Product

For a $m$-dimensional vector space $V^m$ with the common inner product $\boldsymbol{a} \cdot \boldsymbol{b} = \boldsymbol{a}^T \boldsymbol{b} \in \mathbb{R}, \boldsymbol{a}, \boldsymbol{b} \in V^m$, Clifford algebra further defines higher dimensional subspaces constructed by the outer product '$\wedge$'. Subspaces constructed by $k$ vectors using the outer product are denoted as [46], [47]

$$\boldsymbol{a}_1 \wedge \cdots \wedge \boldsymbol{a}_k, \quad \boldsymbol{a}_1, \cdots, \boldsymbol{a}_k \in V^m \tag{1}$$

which is also named as the $k$-blade with grade $k$. The outer product satisfies

$$\boldsymbol{a}_i \wedge \boldsymbol{a}_j = \begin{cases} -\boldsymbol{a}_j \wedge \boldsymbol{a}_i, & i \neq j \\ 0, & i = j \end{cases} \tag{2}$$

$$(\boldsymbol{a}_i \wedge \boldsymbol{a}_j) \wedge \boldsymbol{a}_k = \boldsymbol{a}_i \wedge (\boldsymbol{a}_j \wedge \boldsymbol{a}_k).$$

The geometric product between two Clifford algebras $\boldsymbol{a}, \boldsymbol{b} \in Cl$ is defined by the sum of the inner and outer products [53]

$$\boldsymbol{ab} \triangleq \boldsymbol{a} \cdot \boldsymbol{b} + \boldsymbol{a} \wedge \boldsymbol{b} \tag{3}$$

in which $\boldsymbol{ab} \neq \boldsymbol{ba}$ due to the noncommutativity of the outer product in (2).

Similar to the orthogonal basis for vector spaces, the Clifford algebra can be explicitly defined by the generators $\boldsymbol{e}_1, \cdots, \boldsymbol{e}_n$, and the multiplications between two generators defined by the geometric product [53] satisfy

$$\begin{cases} e_i e_j + e_j e_i = 0, & i \neq j, \ i,j = 1,\cdots,n \\ e_i e_i = 1, \text{ or } e_i e_i = -1, \text{ or } e_i e_i = 0 \end{cases} \quad (4)$$

Taking $e_i e_i = e_j e_j = 1$, $i \neq j$ for instance, we have

$$(e_i e_j)^2 = -e_i e_j e_j e_i = -1 \quad (5)$$

According to the definition in (4), the Clifford algebra can be denoted by $Cl(p,q,r)$, in which $p,q,r$ are the number of generators corresponding to $e_i e_i = 1, -1, 0$, respectively, and $p + q + r = n$. $e_1 e_2 \cdots e_k$ is called a $k$-order multivector, and the Clifford algebra created by $n$ generators has $C_n^k = n!/[k!(n-k)!]$ $k$-order multivectors, with the total number of multivectors being $2^n$. Specifically, the Clifford algebra composed by the even-order and odd-order multivectors is denoted by $Cl^+(p,q,r)$ and $Cl^-(p,q,r)$, respectively.

Based on the generators $\{e_1, e_2, e_3\}$, the general element of the Clifford algebra can be given by

$$Cl = a_0 + a_1 e_1 + a_2 e_2 + a_3 e_3 \\ + a_{12} e_1 e_2 + a_{13} e_1 e_3 + a_{23} e_2 e_3 + a_{123} e_1 e_2 e_3 \quad (6)$$

which contains 0,1,2,3-order multivectors.

**B. Motion Representations by Clifford Algebra**

The Clifford algebras isomorphic[1] to the Special Orthogonal group SO(3) and the Special Euclidean group $SE_{k+2}(3)$ are specifically reviewed herein, respectively. The following part is started by the Clifford algebra corresponding to the familiar quaternion.

*1) Rotation Representation*

The quaternion can be constructed by the even-order Clifford algebra $Cl^+(0,3,0)$, i.e., the three generators satisfy $e_1 e_1 = e_2 e_2 = e_3 e_3 = -1$. The general element is composed by the even-order subspaces of (6) as

$$h = a_0 + a_{12} e_1 e_2 + a_{13} e_1 e_3 + a_{23} e_2 e_3, \ h \in \mathbb{H} \quad (7)$$

According to (4)-(5), the multivectors satisfy

$$(\mathbf{e}_1 \mathbf{e}_2)^2 = (\mathbf{e}_1 \mathbf{e}_3)^2 = (\mathbf{e}_2 \mathbf{e}_3)^2 = (\mathbf{e}_1 \mathbf{e}_2)(\mathbf{e}_1 \mathbf{e}_3)(\mathbf{e}_2 \mathbf{e}_3) = -1 \\ (\mathbf{e}_1 \mathbf{e}_2)(\mathbf{e}_1 \mathbf{e}_3) = \mathbf{e}_2 \mathbf{e}_3 \quad (8)$$

The quaternion usually adopts the familiar definition $e_2 e_3 = i, e_3 e_1 = j, e_1 e_2 = k$, which gives rise to

$$q = a_0 + a_1 i + a_2 j + a_3 k, \ i^2 = j^2 = k^2 = ijk = -1, \\ ij = -ji = k, \ jk = -kj = i, \ ki = -ik = j. \quad (9)$$

The unit quaternion requires $\sqrt{a_0^2 + a_1^2 + a_2^2 + a_3^2} = 1$. Letting $s^2 = a_1^2 + a_2^2 + a_3^2$ and $\mathbf{n} = (a_1 i + a_2 j + a_3 k)/s$, it can be verified that $\mathbf{nn} = -1, \mathbf{n}^3 = -\mathbf{n}, \mathbf{n}^4 = 1, \mathbf{n}^5 = \mathbf{n}, \cdots$. Thus, the quaternion $q = a_0 + s\mathbf{n}$ can be computed by the exponential map

$$q = \exp\left(\frac{\theta}{2}\mathbf{n}\right) = \cos\left(\frac{\theta}{2}\right) + \sin\left(\frac{\theta}{2}\right)\mathbf{n} \quad (10)$$

which can also be denoted as a pair of the scalar and vector parts $q = [\cos(\theta/2), \sin(\theta/2)\mathbf{n}]$.

The multiplication of two quaternions is computed by (9) as

$$q_1 q_2 = [s_1 s_2 - \mathbf{v}_1^T \mathbf{v}_2, s_1 \mathbf{v}_2 + s_2 \mathbf{v}_1 + \mathbf{v}_1 \times \mathbf{v}_2], \\ q_1 = [s_1, \mathbf{v}_1], q_2 = [s_2, \mathbf{v}_2] \quad (11)$$

where '×' is the cross product. It can be found that the cross product is resulted by the noncommutative outer product in (2). Note that the geometric product was explicitly written as the quaternion product ' ∘ ' in [42], [44].

It is already known that the quaternion is locally isomorphic to the matrix group SO(3) with the double cover [54].

*2) Rotation and Translation Representation*

The motion involving with simultaneous rotation and position displacement is universal, and generally represented by the matrix Lie group $SE(3)$ or the dual quaternion.

The dual quaternion can be constructed by the Clifford algebra $Cl^+(0,3,1)$, i.e., the four generators satisfy $e_1 e_1 = e_2 e_2 = e_3 e_3 = -1$, $e_4 e_4 = 0$ [45]-[47]. The general element is composed by the even-order multivectors

$$Cl^+(0,3,1) = a_0 + a_{12} \mathbf{e}_1 \mathbf{e}_2 + a_{13} \mathbf{e}_1 \mathbf{e}_3 + a_{23} \mathbf{e}_2 \mathbf{e}_3 \\ + a_{14} \mathbf{e}_1 \mathbf{e}_4 + a_{24} \mathbf{e}_2 \mathbf{e}_4 + a_{34} \mathbf{e}_3 \mathbf{e}_4 + a_{1234} \mathbf{e}_1 \mathbf{e}_2 \mathbf{e}_3 \mathbf{e}_4 \quad (12)$$

As in (9), the multivectors can be alternatively denoted with the familar symbols

$$\mathbf{e}_2 \mathbf{e}_3 = i, \ \mathbf{e}_3 \mathbf{e}_1 = j, \ \mathbf{e}_1 \mathbf{e}_2 = k, \ \varepsilon = \mathbf{e}_1 \mathbf{e}_2 \mathbf{e}_3 \mathbf{e}_4 \quad (13)$$

where $\varepsilon$ is termed as the dual unit and $i\varepsilon = \varepsilon i, j\varepsilon = \varepsilon j, k\varepsilon = \varepsilon k, \varepsilon^2 = 0$.

The dual quaternion in (12) can be denoted as

$$\hat{q} = a_0 + a_1 i + a_2 j + a_3 k + \varepsilon(a_4 + a_5 i + a_6 j + a_7 k) \\ = \hat{a} + \hat{A}, \ a_0, \cdots, a_7 \in \mathbb{R} \quad (14)$$

in which, $\hat{a} = a_0 + \varepsilon a_4$ is the dual number and $\hat{A} = a_1 i + a_2 j + a_3 k + \varepsilon(a_5 i + a_6 j + a_7 k)$ is the dual vector (pure vector).

Similar to (10), the exponential map for the dual quaternion can be derived as [55]

$$\hat{q} = \exp\left(\frac{\hat{\theta}\hat{\mathbf{n}}}{2}\right) = \cos\left(\frac{\hat{\theta}}{2}\right) + \sin\left(\frac{\hat{\theta}}{2}\right)\hat{\mathbf{n}} \quad (15)$$

where $\hat{\theta} = \theta + \varepsilon d, \hat{\mathbf{n}} = \mathbf{n} + \varepsilon \mathbf{c} \times \mathbf{n}$, $\theta$ is the rotation angle, $\mathbf{n}$ is the rotation axis, $d$ is the distance moving along $\mathbf{n}$, and $\mathbf{c}$ is the vector from the origin of the initial frame to the axis $\mathbf{n}$, perpendicular to $\mathbf{n}$ [43], [55].

---

[1] Given two groups $(G, *)$ and $(H, \odot)$, for all $u$ and $v$ in $G$, if there exists the bijective map $f: G \to H$, and $f(u * v) = f(u) \odot f(v)$, then $G$ is isomorphic to $H$, written as $G \cong H$ [56].

By expanding (15), the dual quaternion can be written in the familiar form as

$$\hat{q} = q + \varepsilon \frac{1}{2} t q \qquad (16)$$

where $q$ is the attitude quaternion, and the position translation $t$ can be denoted by $\theta, c, n$ as given in [43]. Note that $t$ is regarded as the vector quaternion $[0, t]$ in multiplying with $q$.

It can be verified that the dual quaternions with the geometric product as the group operation satisfy the semi-direct product[2], i.e., the sequential operation

$$\begin{aligned}\hat{q}_2 \hat{q}_1 &= \left(q_2 + \varepsilon \frac{1}{2} t_2 q_2\right)\left(q_1 + \varepsilon \frac{1}{2} t_1 q_1\right) \\ &= q_2 q_1 + \varepsilon \frac{1}{2}\left(t_2 + q_2 t_1 q_2^*\right) q_2 q_1\end{aligned} \qquad (17)$$

is equivalent to the single dual quaternion with the attitude $q = q_2 q_1$ and the translation $t = t_2 + q_2 t_1 q_2^*$, where '$*$' denotes the conjugate of a quaternion. Similarly, for the group $SE(3) = SO(3) \ltimes \mathbb{R}^3$ the semi-direct product of two poses is $(R_2, t_2)(R_1, t_1) = (R_2 R_1, t_2 + R_2 t_1)$. It can be easily verified that the dual quaternion is isomorphic to the matrix group $SE(3)$.

*3) Extended Pose Representation*

The extended pose enclosing the attitude, position and velocity is represented by the matrix group $SE_2(3)$ in robotics, also known as the double direct spatial isometries in [23]-[26]. The trident quaternion proposed in [44] also serves the same purpose. In [43], the trident quaternion is defined by the subgroup of the Clifford algebra $Cl^+(0,3,2)$ with $e_1 e_1 = e_2 e_2 = e_3 e_3 = -1$, $e_4 e_4 = 0$, $e_5 e_5 = 0$, and additionally requires $e_4 e_5 = 0$, which ensures the independence of position w.r.t. velocity in representation. The general element of a trident quaternion is composed by the even-order multivectors

$$\begin{aligned}h = &a_0 + a_1 e_1 e_2 + a_2 e_1 e_3 + a_3 e_2 e_3 \\ &+ a_5 e_1 e_4 + a_6 e_2 e_4 + a_7 e_3 e_4 + a_4 e_1 e_2 e_3 e_4 \\ &+ a_9 e_1 e_5 + a_{10} e_2 e_5 + a_{11} e_3 e_5 + a_8 e_1 e_2 e_3 e_5\end{aligned} \qquad (18)$$

Taking $e_2 e_3 = i$, $e_3 e_1 = j$, $e_1 e_2 = k$, $\varepsilon_1 = e_1 e_2 e_3 e_4$ and $\varepsilon_2 = e_1 e_2 e_3 e_5$, we have

$$\begin{aligned}&\varepsilon_1 i = i\varepsilon_1, \varepsilon_1 j = j\varepsilon_1, \varepsilon_1 k = k\varepsilon_1, \varepsilon_2 i = i\varepsilon_2, \\ &\varepsilon_2 j = j\varepsilon_2, \varepsilon_2 k = k\varepsilon_2, \varepsilon_1^2 = \varepsilon_2^2 = \varepsilon_1 \varepsilon_2 = 0\end{aligned} \qquad (19)$$

And, (18) can be accordingly simplified as

$$\breve{q} = q + \varepsilon_1 q' + \varepsilon_2 q'' \qquad (20)$$

where $q, q', q''$ are quaternions.

As in (15), the exponential map of a trident quaternion is given by

$$\breve{q} = \exp\left(\frac{\breve{\theta} \breve{n}}{2}\right) = \cos\left(\frac{\breve{\theta}}{2}\right) + \sin\left(\frac{\breve{\theta}}{2}\right)\breve{n} \qquad (21)$$

---

[2] Given group $G$ and commutable group $H$, and for any $g \in G, h \in H$, the semi-direct product of the pair $(g, h)$ satisfies $(g_2, h_2)(g_1, h_1) = (g_2 g_1, h_2 + g_2 (h_1))$, and it is denoted by $G \ltimes H$.

where $\breve{\theta} = \theta + \varepsilon_1 d_1 + \varepsilon_2 d_2$, $\breve{n} = n + \varepsilon_1 (c_1 \times n) + \varepsilon_2 (c_2 \times n)$, and $c_1, d_1, c_2, d_2$ are the parameters corresponding to two translations. Besides, (21) can also be denoted by two translations as follows.

$$\breve{q} = q + \varepsilon_1 \frac{1}{2} t_1 q + \varepsilon_2 \frac{1}{2} t_2 q \qquad (22)$$

As in (17), it can be readily verified that the trident quaternion also satisfies the semi-direct product, and thus, it is isomorphic to the matrix group $SE_2(3)$. Actually, the exponential maps in (10), (15), (21) provide the mapping from Lie algebras to the corresponding Lie groups.

For small rotation and translations, i.e. $\theta \to 0, t_1, t_2 \to 0$, the Clifford algebra can be approximated by [43]

$$\Delta \breve{q} \approx 1 + \frac{1}{2}(\theta n + \varepsilon_1 t_1 + \varepsilon_2 t_2) \qquad (23)$$

Following the similar rationale, the Clifford algebra isomorphic to group $SE_{k+2}(3)$ can be formulated by directly extending (22). If the $q_O^N$ defines the rotation from frame $O$ to frame $N$, the 3-dimensional vectors $t^O, t^N$ in two frames can be represented in the Clifford algebra

$$\begin{aligned}\breve{q}_O^N = &q_O^N + \varepsilon_1 \frac{1}{2} t_1^O q_O^N + \varepsilon_2 \frac{1}{2} t_2^O q_O^N \\ &+ \varepsilon_3 \frac{1}{2} q_O^N t_3^N + \cdots + \varepsilon_{k+2} \frac{1}{2} q_O^N t_{k+2}^N\end{aligned} \qquad (24)$$

where $\varepsilon_i^2 = \varepsilon_i \varepsilon_j = 0, i, j = 1, \cdots, k+2$.

## III. QUASI-GROUP-AFFINE KINEMATICS FOR INS

The low-cost INS/GNSS integrated navigation system usually estimates the attitude, velocity, position, IMU biases, and lever arms simultaneously. This section first expresses the system states by the Clifford algebra, and then the left- and right-error Clifford algebra errors are formulated to design the quasi-group-affine system.

**A. Kinematics for System States**

As in [42]-[44], [59], the earth frame is selected as the reference frame and the attitude quaternion is also defined as $q_e^b$. To represent the body-frame vectors, i.e., the IMU biases and lever arms, the following Clifford algebra can be defined according to (24)

$$\begin{aligned}\breve{q}_e^b = &q_e^b + \varepsilon_1 \frac{1}{2}\left(C_i^e \dot{r}^i\right) q_e^b + \varepsilon_2 \frac{1}{2}\left(C_i^e r^i\right) q_e^b \\ &+ \varepsilon_3 \frac{1}{2} q_e^b b_g + \varepsilon_4 \frac{1}{2} q_e^b b_a + \varepsilon_5 \frac{1}{2} q_e^b l^b\end{aligned} \qquad (25)$$

where $C_i^e \dot{r}^i = \omega_{ie}^e \times r^e + v^e$ denotes the inertial velocity expressed in the earth frame, $C_i^e r^i = r^e$ is the earth-frame position, $\boldsymbol{v}^e$ is the earth-frame velocity, $\boldsymbol{\omega}_{ie}^e=[0,0,\omega_{ie}^e]^T$, and $\boldsymbol{b}_g, \boldsymbol{b}_a, \boldsymbol{l}^b$ are the gyroscope bias, accelerometer bias and lever arms, respectively.

The left error of (25) is denoted by

$$\Delta \breve{q}_l = \hat{\breve{q}}_e^{b*} \breve{q}_e^b$$
$$= \Delta q_l + \frac{1}{2}\varepsilon_1 \Delta q_l Ad_{q_e^{b*}}\left(\boldsymbol{v}^e - \hat{\boldsymbol{v}}^e\right)$$
$$+ \frac{1}{2}\varepsilon_2 \Delta q_l Ad_{q_e^{b*}}\left(\boldsymbol{r}^e - \hat{\boldsymbol{r}}^e\right) + \frac{1}{2}\varepsilon_3 \Delta q_l \left(\boldsymbol{b}_g - Ad_{\Delta q_l^*}\hat{\boldsymbol{b}}_g\right) \quad (26)$$
$$+ \frac{1}{2}\varepsilon_4 \Delta q_l \left(\boldsymbol{b}_a - Ad_{\Delta q_l^*}\hat{\boldsymbol{b}}_a\right) + \frac{1}{2}\varepsilon_5 \Delta q_l \left(\boldsymbol{l}^b - Ad_{\Delta q_l^*}\hat{\boldsymbol{l}}^b\right)$$

in which $\boldsymbol{v}^e = \boldsymbol{C}_i^e \dot{\boldsymbol{r}}^i$, $\Delta q_l = q_{\hat{b}}^b$, and the adjoint operation of a quaternion is defined as $Ad_q \boldsymbol{x} = q\boldsymbol{x}q^*, \boldsymbol{x} \in \mathbb{R}^3$. For example, the coordinate transformation using the quaternion $q_e^b$ is denoted as $Ad_{q_e^{b*}}\boldsymbol{x}^e = q_e^{b*}\boldsymbol{x}^e q_e^b \triangleq \boldsymbol{C}_e^b \boldsymbol{x}^e$.

The state errors represented in (26) can be rewritten in the vector format as

$$\Delta q_l = \exp(\Delta \boldsymbol{\sigma}_l /2), \quad \Delta \boldsymbol{v}_l^e = \boldsymbol{C}_b^{eT}\left(\boldsymbol{v}^e - \hat{\boldsymbol{v}}^e\right)$$
$$\Delta \boldsymbol{r}_l^e = \boldsymbol{C}_b^{eT}\left(\boldsymbol{r}^e - \hat{\boldsymbol{r}}^e\right), \quad \Delta \boldsymbol{b}_{g,l} = \boldsymbol{b}_g - \boldsymbol{C}_b^{eT}\hat{\boldsymbol{C}}_b^e \hat{\boldsymbol{b}}_g \quad (27)$$
$$\Delta \boldsymbol{b}_{a,l} = \boldsymbol{b}_a - \boldsymbol{C}_b^{eT}\hat{\boldsymbol{C}}_b^e \hat{\boldsymbol{b}}_a, \quad \Delta \boldsymbol{l}_l^b = \boldsymbol{l}^b - \boldsymbol{C}_b^{eT}\hat{\boldsymbol{C}}_b^e \hat{\boldsymbol{l}}^b$$

Similarly, the right error of (25) is denoted by

$$\Delta \breve{q}_r = \breve{q}_e^b \hat{\breve{q}}_e^{b*}$$
$$= \Delta q_r + \frac{1}{2}\varepsilon_1 \Delta q_r \left(Ad_{\Delta q_r^*}\boldsymbol{v}^e - \hat{\boldsymbol{v}}^e\right)$$
$$+ \frac{1}{2}\varepsilon_2 \Delta q_r \left(Ad_{\Delta q_r^*}\boldsymbol{r}^e - \hat{\boldsymbol{r}}^e\right) + \frac{1}{2}\varepsilon_3 \Delta q_r Ad_{\hat{q}_e^b}\left(\boldsymbol{b}_g - \hat{\boldsymbol{b}}_g\right) \quad (28)$$
$$+ \frac{1}{2}\varepsilon_4 \Delta q_r Ad_{\hat{q}_e^b}\left(\boldsymbol{b}_a - \hat{\boldsymbol{b}}_a\right) + \frac{1}{2}\varepsilon_5 \Delta q_r Ad_{\hat{q}_e^b}\left(\boldsymbol{l} - \hat{\boldsymbol{l}}^b\right)$$

in which $\Delta q_r = q_{\hat{e}}^e$.

According to (24), the vector errors represented by (28) are indeed

$$\Delta q_r = \exp(\Delta \boldsymbol{\sigma}_r /2), \quad \Delta \boldsymbol{v}_r^e = \hat{\boldsymbol{C}}_b^e \boldsymbol{C}_b^{eT}\boldsymbol{v}^e - \hat{\boldsymbol{v}}^e$$
$$\Delta \boldsymbol{r}_r^e = \hat{\boldsymbol{C}}_b^e \boldsymbol{C}_b^{eT}\boldsymbol{r}^e - \hat{\boldsymbol{r}}^e, \quad \Delta \boldsymbol{b}_{g,r} = \hat{\boldsymbol{C}}_b^e\left(\boldsymbol{b}_g - \hat{\boldsymbol{b}}_g\right) \quad (29)$$
$$\Delta \boldsymbol{b}_{a,r} = \hat{\boldsymbol{C}}_b^e\left(\boldsymbol{b}_a - \hat{\boldsymbol{b}}_a\right), \quad \Delta \boldsymbol{l}_r^b = \hat{\boldsymbol{C}}_b^e\left(\boldsymbol{l}^b - \hat{\boldsymbol{l}}^b\right)$$

where $\boldsymbol{C}_b^e, \hat{\boldsymbol{C}}_b^e$ are the direction cosine matrices corresponding to $q_e^b, \hat{q}_e^b$, i.e., $Ad_{q_e^b}\boldsymbol{x}^b \triangleq \boldsymbol{C}_b^e \boldsymbol{x}^b, \boldsymbol{x}^b \in \mathbb{R}^3$.

The system kinematics of (25) is given by the differential equation

$$\dot{\breve{q}}_e^b = f\left(\breve{q}_e^b, u_t\right) \triangleq f_{u_t}\left(\breve{q}_e^b\right) \quad (30)$$

in which, $u_t$ denotes the inputs from IMU.

*Remark 1*: In order to make the error kinematics of (30) autonomous, i.e., independent of states, the group-affine condition proposed in [23] requires that

$$\breve{q}_1 f_{u_t}(\breve{q}_2) + f_{u_t}(\breve{q}_1)\breve{q}_2 - \breve{q}_1 f_{u_t}(1)\breve{q}_2 = f_{u_t}(\breve{q}_1 \breve{q}_2) \quad (31)$$

It can be verified that the following system kinematics for the trident quaternion proposed in [42] are group-affine if both IMU biases and the lever arm are not considered.

$$\dot{\breve{q}}_e^b = \left(\breve{q}_e^b \breve{\omega}_{ib}^b - \breve{\omega}_{ie}^e \breve{q}_e^b\right)/2,$$
$$\begin{cases}\breve{\omega}_{ib}^b = \boldsymbol{\omega}_{ib}^b + \varepsilon_1 \boldsymbol{f}^b \\ \breve{\omega}_{ie}^e = \boldsymbol{\omega}_{ie}^e - \varepsilon_1 \boldsymbol{g}^e - \varepsilon_2 \boldsymbol{C}_i^e \dot{\boldsymbol{r}}^i\end{cases} \quad (32)$$

in which $\breve{q}_e^b = q_e^b + \varepsilon_1 \left(\boldsymbol{C}_i^e \dot{\boldsymbol{r}}^i\right)q_e^b/2 + \varepsilon_2 \left(\boldsymbol{C}_i^e \boldsymbol{r}^i\right)q_e^b/2$.

The left and right error kinematics for (32) can be accordingly computed as below

$$\Delta \dot{\breve{q}}_l = \begin{pmatrix}\Delta \breve{q}_l\left(\boldsymbol{\omega}_{ib}^b + \varepsilon_1 \boldsymbol{f}^b\right) \\ -\left(\boldsymbol{\omega}_{ib}^b + \varepsilon_1 \boldsymbol{f}^b\right)\Delta \breve{q}_l\end{pmatrix}\bigg/2 + \varepsilon_2 \Delta q_l',$$
$$\Delta \dot{\breve{q}}_r = \begin{pmatrix}\Delta \breve{q}_r\left(\boldsymbol{\omega}_{ie}^e - \varepsilon_1 \boldsymbol{g}^e\right) \\ -\left(\boldsymbol{\omega}_{ie}^e - \varepsilon_1 \boldsymbol{g}^e\right)\Delta \breve{q}_r\end{pmatrix}\bigg/2 + \varepsilon_2 \Delta q_r' \quad (33)$$

After modeling the body-frame vectors along with the extended pose on the Lie group in (25), it is obvious that system kinematics (30) of (25) are not group-affine and the perfect autonomous error kinematics as in (33) are theoretically unavailable. Therefore, this work terms the kinematics of (25) as quasi-group-affine.

**B. Error Kinematics Selection**

In [42], the trident-quaternion EKFs based on the error kinematics in (33) perform much better than the standard EKF even though the errors of biases and lever arms are defined as additive. However, for low-cost IMU the performance of the "imperfect IEKF" is unsatisfactory, and the convergence region is severely compressed as indicated in [32], [34], [35]. As stated above, the kinematics of (25) is not group-affine due to the interaction of additive biases with the attitude and velocity. Recent work [35] shows that modeling the body-frame vectors on the two-frames group (TFG) further improves the filtering performance in contrast with the imperfect IEKF, but it is only available for the attitude estimation of INS/GNSS system on the abelian group SO(2) (see Theorem 5 of [35]). For the state estimation on group SO(3), it would fail in denoting the derivatives of $\Delta \boldsymbol{b}_{g,l}, \Delta \boldsymbol{b}_{a,l}, \Delta \boldsymbol{l}_l^b$ in (27) by the error states themselves. This is because the differential of $\boldsymbol{C}_b^{eT}\hat{\boldsymbol{C}}_b^e$ generates convoluted nonlinear terms, which are hard to be denoted as the linearized form. In contrast, the derivatives of right errors $\Delta \boldsymbol{b}_{g,r}, \Delta \boldsymbol{b}_{a,r}, \Delta \boldsymbol{l}_r^b$ in (29) can be directly computed by

$$\Delta \dot{\boldsymbol{b}}_{g,r} = \left(\hat{\boldsymbol{C}}_b^e \hat{\boldsymbol{\omega}}_{eb}^b\right) \times \Delta \boldsymbol{b}_{g,r}$$
$$\Delta \dot{\boldsymbol{b}}_{a,r} = \left(\hat{\boldsymbol{C}}_b^e \hat{\boldsymbol{\omega}}_{eb}^b\right) \times \Delta \boldsymbol{b}_{a,r} \quad (34)$$
$$\Delta \dot{\boldsymbol{l}}_r^b = \left(\hat{\boldsymbol{C}}_b^e \hat{\boldsymbol{\omega}}_{eb}^b\right) \times \Delta \boldsymbol{l}_r^b$$

Hence, the right error kinematics of body-frame vectors can be linearized more conveniently comparing with the left counterparts. As shown in [35], the TFG-IEKF achieves higher estimation accuracy than the imperfect IEKF in the low-cost inertial navigation with body-frame observations using the right errors.

Nevertheless, for the INS/GNSS integrated navigation system, the superiority of using the left error in designing the IEKF has been confirmed in [27]-[30], etc. The contradiction between using the left error and modeling the error kinematics of body-frame errors in (34) is hard to be tackled theoretically and still an open problem so far. Therefore, the subsequent section is endeavored to narrow the performance gaps of the filters designed by different state errors.

## IV. CLIFFORD-RQEKF WITH ITERATED UPDATES

According to the right state error in (28), the kinematic models are linearized to design the Clifford-RQEKF method. The characteristics of the models of the standard EKF, LQEKF, RQEKF and Clifford-RQEKF are compared and analyzed. The GNSS velocity measurement model is also linearized, and the iterated filtering technique is introduced to eliminate the remaining linearization errors in the measurement matrices.

### A. Linearization of Error Kinematics and Measurement Models

Based on (28) and (34), the error kinematics for the attitude, transformed velocity, position, biases and lever arms can be derived as

$$\Delta \dot{\breve{q}}_r = \frac{1}{2} \Delta q_r \begin{pmatrix} \Delta \boldsymbol{\omega}_r^e + \varepsilon_1 \left( \Delta \dot{\boldsymbol{r}}_r^e + \Delta \boldsymbol{\omega}_r^e \times \Delta \boldsymbol{r}_r^e \right) \\ + \varepsilon_2 \left( \Delta \dot{\boldsymbol{v}}_r^e + \Delta \boldsymbol{\omega}_r^e \times \Delta \boldsymbol{v}_r^e \right) \\ + \varepsilon_3 \left( \left( \hat{\boldsymbol{C}}_b^e \hat{\boldsymbol{\omega}}_{eb}^b \right) \times \Delta \boldsymbol{b}_{g,r} + \hat{\boldsymbol{C}}_b^e \boldsymbol{n}_{ba} \right) \\ + \varepsilon_4 \left( \left( \hat{\boldsymbol{C}}_b^e \hat{\boldsymbol{\omega}}_{eb}^b \right) \times \Delta \boldsymbol{b}_{a,r} + \hat{\boldsymbol{C}}_b^e \boldsymbol{n}_{bg} \right) \\ + \varepsilon_5 \left( \left( \hat{\boldsymbol{C}}_b^e \hat{\boldsymbol{\omega}}_{eb}^b \right) \times \Delta \boldsymbol{l}_r^b + \hat{\boldsymbol{C}}_b^e \boldsymbol{n}_l \right) \end{pmatrix} \quad (35)$$

in which, $\boldsymbol{n}_{bg}, \boldsymbol{n}_{ba}, \boldsymbol{n}_l$ are the noises corresponding to the IMU biases and lever arms, and the derivative of the attitude error is computed by

$$\begin{aligned} \Delta \dot{q}_r &= q_e^b \boldsymbol{\omega}_{eb}^b \hat{q}_e^{b*} - q_e^b \hat{\boldsymbol{\omega}}_{eb}^b \hat{q}_e^{b*} \\ &= \Delta q_r Ad_{\hat{q}_e^b} \left( \boldsymbol{\omega}_{eb}^b - \hat{\boldsymbol{\omega}}_{eb}^b \right) \\ &= \Delta q_r \Delta \boldsymbol{\omega}_r^e \end{aligned} \quad (36)$$

For small state errors, the Clifford algebra error in (28) can be approximated to the first order as in (23) [42], [43].

$$\Delta \breve{q}_r \approx 1 + \frac{1}{2} \Delta \boldsymbol{\sigma}_r + \varepsilon_1 \frac{1}{2} \Delta \boldsymbol{\sigma}_v + \frac{1}{2} \varepsilon_2 \Delta \boldsymbol{\sigma}_p \\ + \varepsilon_3 \frac{1}{2} \Delta \boldsymbol{\sigma}_{b_g} + \frac{1}{2} \varepsilon_4 \Delta \boldsymbol{\sigma}_{b_a} + \frac{1}{2} \varepsilon_5 \Delta \boldsymbol{\sigma}_l \quad (37)$$

in which $\Delta \boldsymbol{\sigma}_r, \Delta \boldsymbol{\sigma}_v, \Delta \boldsymbol{\sigma}_p, \Delta \boldsymbol{\sigma}_{b_g}, \Delta \boldsymbol{\sigma}_{b_a}, \Delta \boldsymbol{\sigma}_l \in \mathbb{R}^3$ indeed denote the components of the Lie algebra of $\Delta \breve{q}_r$.

Substituting (37) into (35) and omitting the higher-order terms in the right-hand side of (35), the 18-dimensional differential equation is obtained.

$$\begin{aligned} \Delta \dot{\boldsymbol{\sigma}}_r &\approx \Delta \boldsymbol{\omega}_r^e = Ad_{\hat{q}_e^{b*}} \left( \boldsymbol{\omega}_{eb}^b - \hat{\boldsymbol{\omega}}_{eb}^b \right) \\ \Delta \dot{\boldsymbol{\sigma}}_v &\approx \Delta \dot{\boldsymbol{v}}_r^e = Ad_{\Delta q_r *} \dot{\boldsymbol{v}}^e - \dot{\hat{\boldsymbol{v}}}^e + Ad_{\Delta q_r} \boldsymbol{v}^e \times \Delta \boldsymbol{\omega}_r^e \\ \Delta \dot{\boldsymbol{\sigma}}_p &\approx \Delta \dot{\boldsymbol{r}}_r^e = Ad_{\Delta q_r *} \dot{\boldsymbol{r}}^e - \dot{\hat{\boldsymbol{r}}}^e + Ad_{\Delta q_r} \boldsymbol{r}^e \times \Delta \boldsymbol{\omega}_r^e \\ \Delta \dot{\boldsymbol{\sigma}}_{b_g} &\approx \Delta \dot{\boldsymbol{b}}_{g,r} = \left( \hat{\boldsymbol{C}}_b^e \hat{\boldsymbol{\omega}}_{eb}^b \right) \times \Delta \boldsymbol{b}_{g,r} + \hat{\boldsymbol{C}}_b^e \boldsymbol{n}_{ba} \\ \Delta \dot{\boldsymbol{\sigma}}_{b_a} &\approx \Delta \dot{\boldsymbol{b}}_{a,r} = \left( \hat{\boldsymbol{C}}_b^e \hat{\boldsymbol{\omega}}_{eb}^b \right) \times \Delta \boldsymbol{b}_{a,r} + \hat{\boldsymbol{C}}_b^e \boldsymbol{n}_{bg} \\ \Delta \dot{\boldsymbol{\sigma}}_l &\approx \Delta \dot{\boldsymbol{l}}_r^b = \left( \hat{\boldsymbol{C}}_b^e \hat{\boldsymbol{\omega}}_{eb}^b \right) \times \Delta \boldsymbol{l}_r^b + \hat{\boldsymbol{C}}_b^e \boldsymbol{n}_l \end{aligned} \quad (38)$$

in which $\boldsymbol{\omega}_{eb}^b = \boldsymbol{\omega}_{ib}^b - Ad_{\hat{q}_e^{b*}} \boldsymbol{\omega}_{ie}^e$ and the kinematics of the transformed velocity and position are given as [27], [44]

$$\begin{aligned} \dot{\boldsymbol{v}}^e &= Ad_{q_e^b} \boldsymbol{f}^b - \boldsymbol{\omega}_{ie}^e \times \boldsymbol{v}^e + \boldsymbol{g}^e \\ \dot{\boldsymbol{r}}^e &= \boldsymbol{v}^e - \boldsymbol{\omega}_{ie}^e \times \boldsymbol{r}^e \end{aligned} \quad (39)$$

Substituting (39) into (38) and omitting higher-order terms, the Jacobian matrix for the error states can be obtained for the linearized system.

$$\Delta \boldsymbol{x}_r = \begin{bmatrix} \Delta \boldsymbol{\sigma}_r^T & \Delta \boldsymbol{\sigma}_v^T & \Delta \boldsymbol{\sigma}_p^T & \Delta \boldsymbol{\sigma}_{b_g}^T & \Delta \boldsymbol{\sigma}_{b_a}^T & \Delta \boldsymbol{\sigma}_l^T \end{bmatrix}^T \quad (40)$$

$$\Delta \dot{\boldsymbol{x}}_r = \boldsymbol{F}_r \Delta \boldsymbol{x}_r + \boldsymbol{G}_r \boldsymbol{w} \quad (41)$$

$$\boldsymbol{w} = \begin{bmatrix} \boldsymbol{w}_{grw}^T & \boldsymbol{w}_{arw}^T & \boldsymbol{n}_g^T & \boldsymbol{n}_a^T & \boldsymbol{n}_l^T \end{bmatrix}^T \quad (42)$$

$$\boldsymbol{F}_r = \begin{bmatrix} -\boldsymbol{\omega}_{ie}^e \times & \boldsymbol{0} & \boldsymbol{0} & -\boldsymbol{I} & \boldsymbol{0} & \boldsymbol{0} \\ \boldsymbol{g}^e \times & -\boldsymbol{\omega}_{ie}^e \times & \boldsymbol{0} & -\hat{\boldsymbol{v}}^e \times & -\boldsymbol{I} & \boldsymbol{0} \\ \boldsymbol{0} & \boldsymbol{I} & -\boldsymbol{\omega}_{ie}^e \times & -\hat{\boldsymbol{r}}^e \times & \boldsymbol{0} & \boldsymbol{0} \\ \boldsymbol{0} & \boldsymbol{0} & \boldsymbol{0} & \boldsymbol{D} & \boldsymbol{0} & \boldsymbol{0} \\ \boldsymbol{0} & \boldsymbol{0} & \boldsymbol{0} & \boldsymbol{0} & \boldsymbol{D} & \boldsymbol{0} \\ \boldsymbol{0} & \boldsymbol{0} & \boldsymbol{0} & \boldsymbol{0} & \boldsymbol{0} & \boldsymbol{D} \end{bmatrix} \quad (43)$$

in which $\boldsymbol{D} = \left( \hat{\boldsymbol{C}}_b^e \hat{\boldsymbol{\omega}}_{eb}^b \right) \times$, $\boldsymbol{0}$ is the 3×3 zero matrix, $\boldsymbol{I}$ is the 3×3 identity matrix, $\boldsymbol{w}_{grw}$, $\boldsymbol{w}_{arw}$ are the random walks of gyroscopes and accelerometers, respectively.

$$\boldsymbol{G}_r = \begin{bmatrix} -\hat{\boldsymbol{C}}_b^e & \boldsymbol{0} & \boldsymbol{0} & \boldsymbol{0} & \boldsymbol{0} \\ -\hat{\boldsymbol{v}}^e \times \hat{\boldsymbol{C}}_b^e & -\hat{\boldsymbol{C}}_b^e & \boldsymbol{0} & \boldsymbol{0} & \boldsymbol{0} \\ -\hat{\boldsymbol{r}}^e \times \hat{\boldsymbol{C}}_b^e & \boldsymbol{0} & \boldsymbol{0} & \boldsymbol{0} & \boldsymbol{0} \\ \boldsymbol{0} & \boldsymbol{0} & \hat{\boldsymbol{C}}_b^e & \boldsymbol{0} & \boldsymbol{0} \\ \boldsymbol{0} & \boldsymbol{0} & \boldsymbol{0} & \hat{\boldsymbol{C}}_b^e & \boldsymbol{0} \\ \boldsymbol{0} & \boldsymbol{0} & \boldsymbol{0} & \boldsymbol{0} & \hat{\boldsymbol{C}}_b^e \end{bmatrix} \quad (44)$$

The detailed derivations for (41)-(44) are given in the Appendix A.

The GNSS velocity measurement in the earth frame is

$$\boldsymbol{y} = \boldsymbol{v}^e + \boldsymbol{C}_b^e \boldsymbol{\omega}_{eb}^b \times \boldsymbol{l}^b \quad (45)$$

The innovation of (45) can be approximated by

$$\begin{aligned}
& y - \hat{y} \\
& = v^e + C_b^e \omega_{eb}^b \times l^b - \hat{v}^e - \hat{C}_b^e \omega_{eb}^b \times \hat{l}^b \\
& \approx v^e - \omega_{ie}^e \times r^e - (\hat{v}^e - \omega_{ie}^e \times \hat{r}^e) \\
& \quad + (I + \Delta\sigma_r \times)\hat{C}_b^e \omega_{eb}^b \times (\hat{l}^b + \delta l^b) - \hat{C}_b^e \omega_{eb}^b \times \hat{l}^b \\
& = v^e - \hat{v}^e - \omega_{ie}^e \times (r^e - \hat{r}^e) - (\hat{C}_b^e \omega_{eb}^b \times \hat{l}^b) \times \Delta\sigma_r \\
& \quad + \hat{C}_b^e \omega_{eb}^b \times \delta l^b
\end{aligned} \tag{46}$$

Since the right error in (29) can be approximated to the first order as $\Delta v_r^e \approx v^e - \hat{v}^e - \Delta\sigma_r \times v^e$, $\Delta r_r^e \approx r^e - \hat{r}^e - \Delta\sigma_r \times r^e$, the measurement matrix is accordingly computed by

$$\begin{aligned}
& H_r = \begin{bmatrix} J_{\Delta\sigma} & I & -\omega_{ie}^e \times & 0 & 0 & (\hat{C}_b^e \omega_{eb}^b) \times \end{bmatrix}, \\
& J_{\Delta\sigma} = -v^e \times + \omega_{ie}^e \times r^e \times - (\hat{C}_b^e \omega_{eb}^b \times \hat{l}^b) \times
\end{aligned} \tag{47}$$

Obviously, the measurement matrix is unavoidably affected by the attitude estimation error especially at the beginning of the filtering process.

In the implementation of the Clifford-RQEKF, the initial covariance matrix for the error states can be computed by approximating the nonlinear errors in (29) to the first order as

$$P_{Clifford} = J P_{add} J^T,$$

$$J = \begin{bmatrix} I & 0 & 0 & 0 & 0 & 0 \\ \hat{v}^e \times & I & 0 & 0 & 0 & 0 \\ \hat{r}^e \times & 0 & I & 0 & 0 & 0 \\ 0 & 0 & 0 & \hat{C}_b^e & 0 & 0 \\ 0 & 0 & 0 & 0 & \hat{C}_b^e & 0 \\ 0 & 0 & 0 & 0 & 0 & \hat{C}_b^e \end{bmatrix} \tag{48}$$

where $P_{add}$ denotes the covariance matrix for additive state errors.

For easy comparison, the Jacobian matrices for the standard EKF and the left/right trident-quaternion EKF (LQEKF/RQEKF) derived with (33) and additive body-frame vector errors, i.e., the errors evolving on the group $SE_2(3) \times \mathbb{R}^9$, are also given as below.

$$F_{EKF} = \begin{bmatrix} -\omega_{ie}^e \times & 0 & 0 & -\hat{C}_b^e & 0 & 0 \\ (\hat{C}_b^e \hat{f}^b) \times & -2\omega_{ie}^e \times & 0 & 0 & \hat{C}_b^e & 0 \\ 0 & I & 0 & 0 & 0 & 0 \\ 0 & 0 & 0 & 0 & 0 & 0 \\ 0 & 0 & 0 & 0 & 0 & 0 \\ 0 & 0 & 0 & 0 & 0 & 0 \end{bmatrix} \tag{49}$$

$$G_{EKF} = \begin{bmatrix} -\hat{C}_b^e & 0 & 0 & 0 & 0 \\ 0 & \hat{C}_b^e & 0 & 0 & 0 \\ 0 & 0 & 0 & 0 & 0 \\ 0 & 0 & I & 0 & 0 \\ 0 & 0 & 0 & I & 0 \\ 0 & 0 & 0 & 0 & I \end{bmatrix} \tag{50}$$

$$\begin{aligned}
& H_{EKF} = \begin{bmatrix} J_{\Delta\sigma} & I & 0 & 0 & 0 & \hat{C}_b^e \omega_{eb}^b \times \end{bmatrix}, \\
& J_{\Delta\sigma} = (\hat{C}_b^e \omega_{eb}^b \times \hat{l}^b) \times
\end{aligned} \tag{51}$$

$$F_{LQ} = \begin{bmatrix} -\hat{\omega}_{ib}^b \times & 0 & 0 & I & 0 & 0 \\ -\hat{f}^b & -\hat{\omega}_{ib}^b \times & 0 & 0 & I & 0 \\ 0 & I & -\hat{\omega}_{ib}^b \times & 0 & 0 & 0 \\ 0 & 0 & 0 & 0 & 0 & 0 \\ 0 & 0 & 0 & 0 & 0 & 0 \\ 0 & 0 & 0 & 0 & 0 & 0 \end{bmatrix} \tag{52}$$

$$G_{LQ} = \begin{bmatrix} -I & 0 & 0 & 0 & 0 \\ 0 & -I & 0 & 0 & 0 \\ 0 & 0 & 0 & 0 & 0 \\ 0 & 0 & I & 0 & 0 \\ 0 & 0 & 0 & I & 0 \\ 0 & 0 & 0 & 0 & I \end{bmatrix} \tag{53}$$

$$\begin{aligned}
& H_{LQ} = \begin{bmatrix} J_{\Delta\sigma} & \hat{C}_b^e & -\omega_{ie}^e \times \hat{C}_b^e & 0 & 0 & \hat{C}_b^e \omega_{eb}^b \times \end{bmatrix}, \\
& J_{\Delta\sigma} = -\hat{C}_b^e (\omega_{eb}^b \times \hat{l}^b) \times
\end{aligned} \tag{54}$$

$$F_{RQ} = \begin{bmatrix} -\omega_{ie}^e \times & 0 & 0 & -\hat{C}_b^e & 0 & 0 \\ g^e \times & -\omega_{ie}^e \times & 0 & -\hat{v}^e \times \hat{C}_b^e & -\hat{C}_b^e & 0 \\ 0 & I & -\omega_{ie}^e \times & -\hat{r}^e \times \hat{C}_b^e & 0 & 0 \\ 0 & 0 & 0 & 0 & 0 & 0 \\ 0 & 0 & 0 & 0 & 0 & 0 \\ 0 & 0 & 0 & 0 & 0 & 0 \end{bmatrix} \tag{55}$$

$$G_{RQ} = \begin{bmatrix} -\hat{C}_b^e & 0 & 0 & 0 & 0 \\ -\hat{v}^e \times \hat{C}_b^e & -\hat{C}_b^e & 0 & 0 & 0 \\ -\hat{r}^e \times \hat{C}_b^e & 0 & 0 & 0 & 0 \\ 0 & 0 & I & 0 & 0 \\ 0 & 0 & 0 & I & 0 \\ 0 & 0 & 0 & 0 & I \end{bmatrix}. \tag{56}$$

$$\begin{aligned}
& H_{RQ} = \begin{bmatrix} J_{\Delta\sigma} & I & -\omega_{ie}^e \times & 0 & 0 & \hat{C}_b^e \omega_{eb}^b \times \end{bmatrix}, \\
& J_{\Delta\sigma} = -v^e \times + \omega_{ie}^e \times r^e \times - (\hat{C}_b^e \omega_{eb}^b \times \hat{l}^b) \times
\end{aligned} \tag{57}$$

Note that the IMU measurements are denoted by $\hat{\omega}_{ib}^b = \tilde{\omega}_{ib}^b - \hat{b}_g$, $\hat{f}^b = \tilde{f}^b - \hat{b}_a$ in (52), which reveals that the LQEKF might be vulnerable to significant IMU biases. As for the process models in (43), (49), (52) and (55), the core upper part of (43) related with the navigation states is perfect with known initial position and velocity, while the evolution of the bias and lever arm errors is primarily affected by the attitude error. The standard EKF is mainly affected by the attitude error, and the error of (52) comes from the error of biases. Comparing (55) with (43), the upper-right components of (55) are affected by the attitude error. In a nutshell, the state dependency is shifted from the components related with the navigation states in RQEKF to the body-frame vectors in Clifford-RQEKF, serving the similar purpose as TFG-IEKF [35]. Since the linearized models in (43)-(44) are still partially state-dependent, this is why the kinematic model of (25) is termed as quasi-group-affine. For the measurement models in (47), (51), (54) and (57), all of them are affected by the initial attitude error.

### B. Iterated Extended Kalman Filtering

To eliminate the linearization errors in the measurement models, the iterated filtering technique is applied to the measurement updates, and for instance the iterated Clifford-RQEKF is accordingly termed as Clifford-RQEKF-Iter. The updating scheme of the iterated Kalman filter is essentially a Gauss-Newton method in approximating the maximum likelihood estimation [48], whose effectiveness has been confirmed in ROVIO [50], Fast-Lio [51], Jung [52] etc. The iterated filtering improves the accuracy of the linearization point for the measurement model using the following update process [49].

$$\begin{aligned} \boldsymbol{\eta}_{i+1} &= \hat{\boldsymbol{x}}_{k+1|k} + \boldsymbol{K}(\boldsymbol{\eta}_i)\left[\boldsymbol{y} - \boldsymbol{h}(\boldsymbol{\eta}_i) - \boldsymbol{H}(\boldsymbol{\eta}_i)(\hat{\boldsymbol{x}}_{k+1|k} - \boldsymbol{\eta}_i)\right], \\ \boldsymbol{K} &= \boldsymbol{P}_{k+1|k}\boldsymbol{H}(\boldsymbol{\eta}_i)^T\left(\boldsymbol{H}(\boldsymbol{\eta}_i)\boldsymbol{P}_{k+1|k}\boldsymbol{H}(\boldsymbol{\eta}_i)^T + \boldsymbol{R}\right)^{-1}, \quad (58)\\ \boldsymbol{\eta}_1 &= \hat{\boldsymbol{x}}_{k+1|k} \end{aligned}$$

in which $i = 1,\ldots,L$, and $\hat{\boldsymbol{x}}_{k+1|k}, \boldsymbol{P}_{k+1|k}$ are the propagated state and covariance, respectively. The iteration is typically finalized with the maximal iteration number $L$ or certain threshold on the magnitude of $\|\boldsymbol{\eta}_{i+1} - \boldsymbol{\eta}_i\|$.

For the indirect Kalman filter widely used in the inertial navigation, the propagated error state is reset as a zero vector, and thus the iterations can be modified as

$$\begin{aligned} \delta\boldsymbol{\eta}_{i+1} &= \boldsymbol{0} + \boldsymbol{K}(\boldsymbol{\eta}_i)\left[\boldsymbol{y} - \boldsymbol{h}(\boldsymbol{\eta}_i) - \boldsymbol{H}(\boldsymbol{\eta}_i)(\boldsymbol{0} - \delta\boldsymbol{\eta}_i)\right], \\ \delta\boldsymbol{\eta}_1 &= \boldsymbol{0}, \boldsymbol{\eta}_1 = \hat{\boldsymbol{x}}_{k+1|k}, \boldsymbol{\eta}_{i+1} = \hat{\boldsymbol{x}}_{k+1|k} \otimes \delta\boldsymbol{\eta}_{i+1} \end{aligned} \quad (59)$$

where $\Delta\boldsymbol{x}_r = \delta\boldsymbol{\eta}_{i+1}$ and $\boldsymbol{K}$ is computed as in (58). The operator '$\otimes$' is defined according to the right errors defined in (29), i.e.,

$$\begin{aligned} q_e^b &= \exp(\Delta\boldsymbol{\sigma}_r/2)\hat{q}_e^b, & \boldsymbol{v}^e &= \boldsymbol{C}_b^e\hat{\boldsymbol{C}}_b^{eT}\left(\Delta\boldsymbol{v}_r^e + \hat{\boldsymbol{v}}^e\right) \\ \boldsymbol{r}^e &= \boldsymbol{C}_b^e\hat{\boldsymbol{C}}_b^{eT}\left(\Delta\boldsymbol{r}_r^e + \hat{\boldsymbol{r}}^e\right), & \boldsymbol{b}_g &= \hat{\boldsymbol{b}}_g + \hat{\boldsymbol{C}}_b^{eT}\Delta\boldsymbol{b}_{g,r} \quad (60) \\ \boldsymbol{b}_a &= \hat{\boldsymbol{b}}_a + \hat{\boldsymbol{C}}_b^{eT}\Delta\boldsymbol{b}_{a,r} & \boldsymbol{l}^b &= \hat{\boldsymbol{l}}^b + \hat{\boldsymbol{C}}_b^{eT}\Delta\boldsymbol{l}_r^b \end{aligned}$$

in which, the state errors in the updates can be computed by the approach given in Appendix B.

Since the attitude is the trickiest to be estimated, the iteration process could be terminated if the magnitude of $\|\Delta\boldsymbol{\sigma}_{r,k+1} - \Delta\boldsymbol{\sigma}_{r,k}\|$ is smaller than a threshold. The iterative process ameliorates the accuracy of the updated states, and the accuracy of the subsequent propagations of states and covariance can be accordingly improved.

## V. SIMULATIONS AND EXPERIMENTS

Simulated and experimental data of the vehicle-based navigation system are used to evaluate the performances of the iterated versions of EKF, LQEKF, RQEKF and Clifford-RQEKF under unknown initial attitude. The maximal iteration number and the terminating threshold are set as 20 and 0.01 deg, respectively.

### A. Numerical Simulations

The car-mounted low-cost INS/GNSS integrated navigation system is simulated herein. The specifications for the simulated low-cost MEMS IMU are listed in Table II. In the simulations, the vehicle moves with uniform speed most of the time, and performs accelerations, decelerations and turnings periodically. The velocity and trajectory of the vehicle is illustrated in Fig. 1. The lever arms between IMU and the GNSS receiver is set as [0.5 0.8 0.3] m. The standard deviation of the GNSS velocity error in the earth frame is 0.1 m/s. This simulation mimics the scenarios where only the 2-dimensional motions of the typical land vehicle are considered as in [60]. Therefore, the horizontal trajectory of Fig. 1 also shows the heading variations of the vehicle.

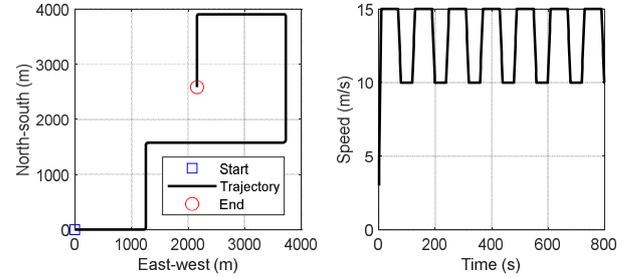

Fig. 1. The simulated trajectory and speed of the vehicle.

To test the algorithms' convergence ability without the prior knowledge of the attitude, the initial attitude error of the vehicle is set to [60, 180, 60] deg for the roll, heading and pitch. The initial standard deviation of the earth-frame attitude error is set as [180, 180, 180] deg to emulate the scenarios where the prior attitude is unavailable. In this work, the EKF initialized with the true attitude, i.e., the ideally initialized EKF (Ideal-EKF), is also implemented as the normal filtering performance. The attitude estimation errors of the Ideal-EKF, EKF-Iter, LQEKF-Iter, RQEKF-Iter and Clifford-RQEKF-Iter are shown in Fig. 2, and only the EKF-Iter cannot converge in a short time, which means that the attitude error contained in (49) is still dominant for the EKF. The heading errors and the corresponding 3-sigma bounds are compared in Fig. 3, which indicate the superior filtering consistency of the proposed methods over EKF-Iter. Specifically, the attitude covariance of EKF-Iter quickly become very small (i.e., $10^{-4}$) and too optimistic just after several updates, refusing further attitude corrections from new measurements. In contrast, the Clifford-RQEKF-Iter, LQEKF-Iter and RQEKF-Iter can converge fast with similar performances.

The estimated gyroscope and accelerometer biases are also compared in Figs. 4-5, in which the accelerometer biases along the forward and lateral directions converge slower than the upward direction due to the limited maneuvers in the simulation. The root mean square of heading errors (RMSE) over specific intervals are compared in Table III (left columns), in which the periods of 1~10 s, 10~50 s and 200~600 s are selected to demonstrate the transient, asymptotic and stabilized phases, respectively. It is shown that the LQEKF-Iter and Clifford-RQEKF-Iter converge to the similar accuracy of the Ideal-EKF in about 20 s, while the RQEKF-Iter converges slightly slower. Note that the rate of convergence is also determined by the vehicle's maneuvers essentially affecting the state observability. As shown in Fig. 6, the estimation accuracy of the lever arms using the velocity-only measurements is not satisfactory, since the accurate estimation of lever arms requires enough angular maneuvers and high-accuracy position measurements as studied in [61]. It might be argued that the current simulation does not well mimic the real vehicle motion dynamics; however, the estimation performances of the proposed methods can be evaluated by comparing with the Ideal-EKF.

TABLE II
**Specifications of the low-cost IMU in the simulations**

| Parameter | Gyroscope | Accelerometer |
|---|---|---|
| Bias | [1000, 500, 800] deg/h | [15, 5, 8] mg |
| Biases instability | 10 deg/h | 20 µg |
| Random walks | 0.2 deg/√h | 200 µg/√Hz |

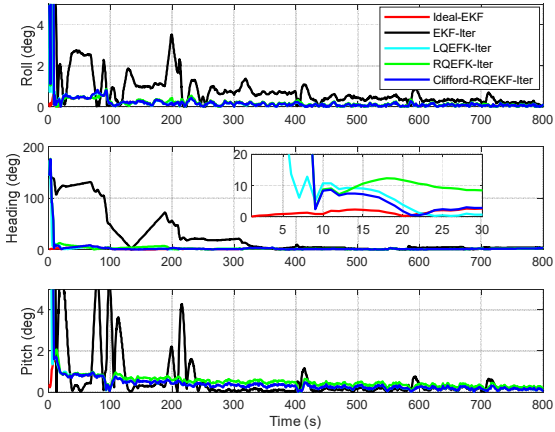

Fig. 2. The attitude estimation errors.

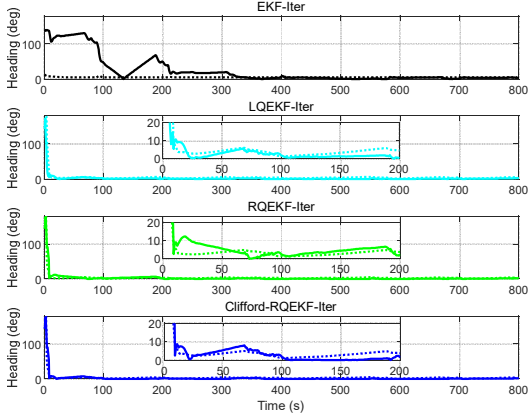

Fig. 3. The heading estimation errors and the corresponding 3-sigma bounds (solid lines: errors, dotted lines: 3-sigma bounds).

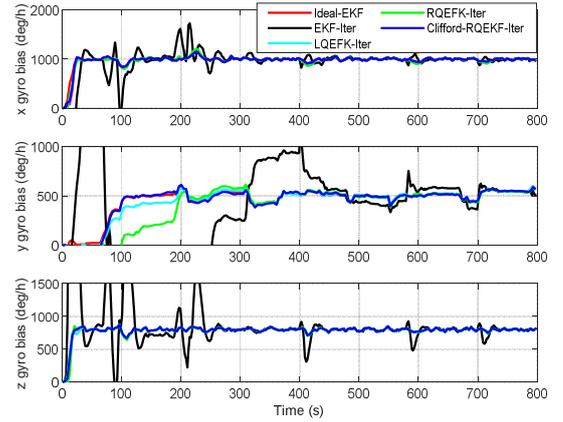

Fig. 4. The estimated gyroscope biases.

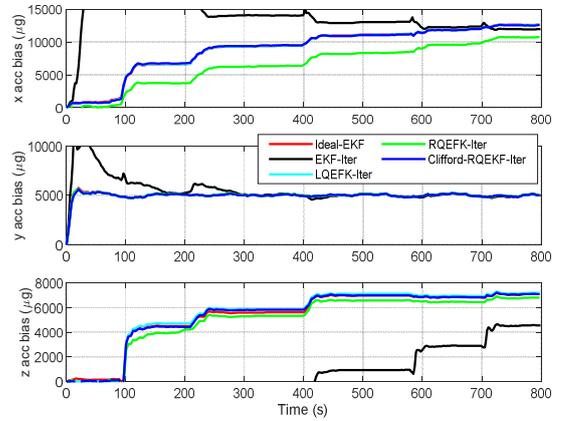

Fig. 5. The estimated accelerometer biases.

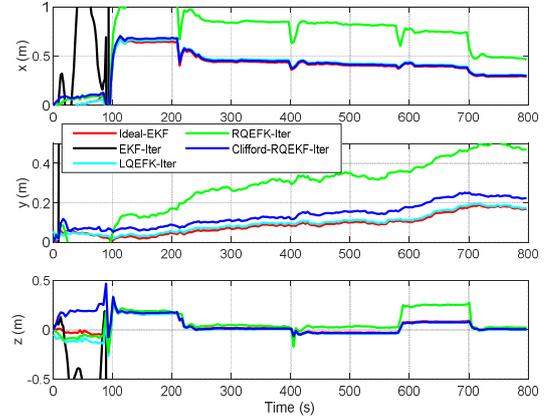

Fig. 6. The estimated lever arms.

TABLE III
**The RMSE of heading errors (deg) in the simulations**

| Methods | 1~10 s | | 10~50 s | | 200~600 s | |
|---|---|---|---|---|---|---|
| Ideal-EKF | 0.88 | 2.27 | 3.05 | 7.29 | 0.98 | 1.35 |
| EKF-Iter | 138 | 129 | 122 | 107 | 12.0 | 100 |
| LQEKF-Iter | 88.9 | 93.9 | 4.64 | 95.4 | 0.95 | 70.3 |
| RQEKF-Iter | 110 | 135 | 8.65 | 11.1 | 1.08 | 4.48 |
| Clifford-RQEKF-Iter | 110 | 108 | 4.64 | 9.62 | 1.11 | 1.13 |

The iteration numbers in Fig. 7 indicate that mostly 2 iterations are required once the attitude successfully converges in about 20 s. This result is consistent with the previous conclusion in [49] w.r.t. the number of iterations necessitated for nonlinear measurement models. As for the EKF-Iter, however, the attitude corrections quickly become very small after several updates due to the false observability problem, and thus only 2 iterations are performed to reach the threshold on the magnitude of the corrections. This phenomenon reveals that the iterated updates are futile for the EKF under large attitude errors.

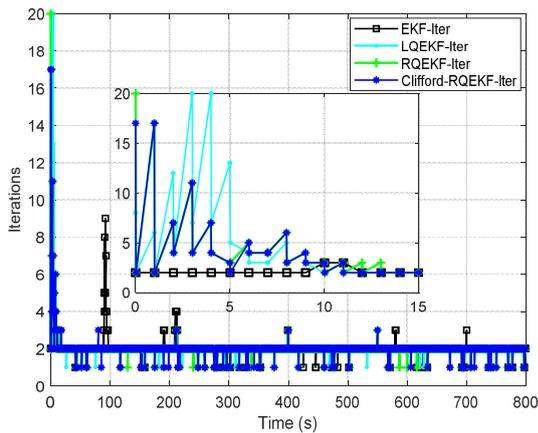

Fig. 7. The number of iterations for each update.

Subsequently, larger IMU biases such as [2000, 2000, 2000] deg/h and [80, 60, 50] mg are considered in the simulations. As shown in Figs. 8-10, the Clifford-RQEKF-Iter still achieves accurate and fast estimations of the attitude and biases, while the LQEKF-Iter and RQEKF-Iter cannot converge, which means that large biases cannot be well-estimated by using the models in (52) and (55). The RMSEs of the heading errors over specific intervals are also compared in the right columns of Table III. Fig. 11 further indicates that only the Clifford-RQEKF-Iter can achieve the stable estimation of lever arms, while the LQEKF-Iter and RQEKF-Iter deviate to abnormally large values.

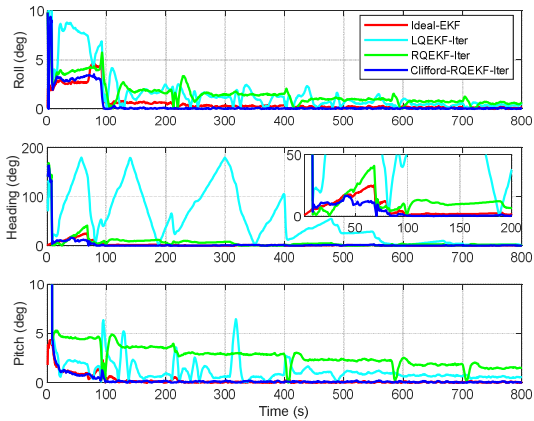

Fig. 8. Attitude estimation errors under intentionally enlarged biases.

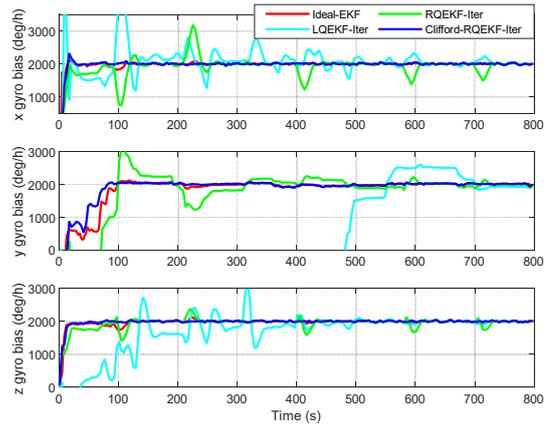

Fig. 9. The estimated gyroscope biases (intentionally enlarged).

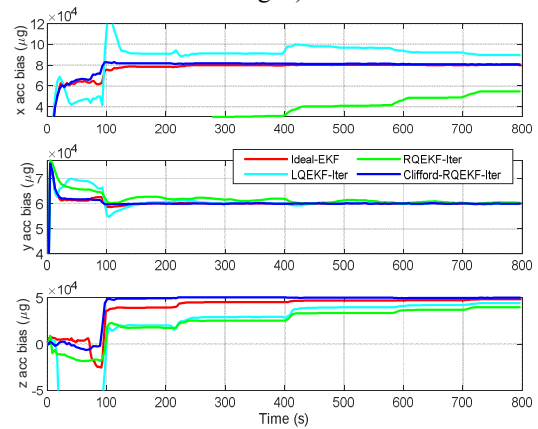

Fig. 10. The estimated accelerometer biases (intentionally enlarged).

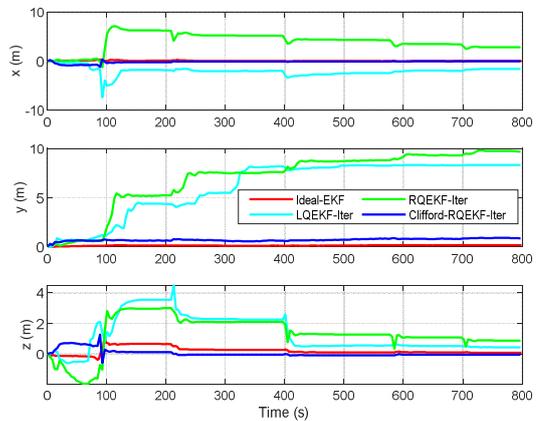

Fig. 11. The estimated lever arms under intentionally enlarged IMU biases.

Note that the EKF, LQEKF, RQEKF and Clifford-RQEKF converge very slowly or even diverge under arbitrarily large initial attitude errors and their results are not provided for brevity. As compared in [57], the performances of advanced invariant/equivariant filters in the transient phase are similar under the attitude errors of only 20 deg in standard deviations. Hence, the iterated filtering technique, as shown herein, is promising to significantly enlarge the convergence regions of the future advanced filters.

## B. Tests by Experimental Data

The land vehicle data collected by a consumer-grade INS/GNSS integrated navigation system [58] are exploited to assess the effectiveness of the proposed method. The specifications of the Bosch BMI088 IMU are shown in Table IV, and the Tersus David30 multi-frequency GNSS receiver collected the pseudorange, doppler, and carrier-phase measurements. The GNSS receiver works in the Real-Time-Kinematic (RTK) mode and the ground truth is obtained by post-processing the data collected with the fiber-optic IMU and the GNSS raw data. The lever arms from the IMU to the GNSS antenna are about [0.28, 0.05, -0.36] m. The standard deviation of the velocity measurement is about 0.1 m/s in the earth frame. The tested trajectory in the Shanghai city along with the vehicle's forward speed are shown in Fig. 12. The GNSS measurements are available in most places nearby the river, but interrupted by the tall buildings occasionally.

**TABLE IV**
**Specifications of the Bosch BMI088 IMU**

| Parameter | Gyroscope | Accelerometer |
|---|---|---|
| Bias | 1deg/s | 20mg |
| Biases instability | 5deg/h | 80μg |
| Random walks | 0.35deg/√h | 200μg/√Hz |

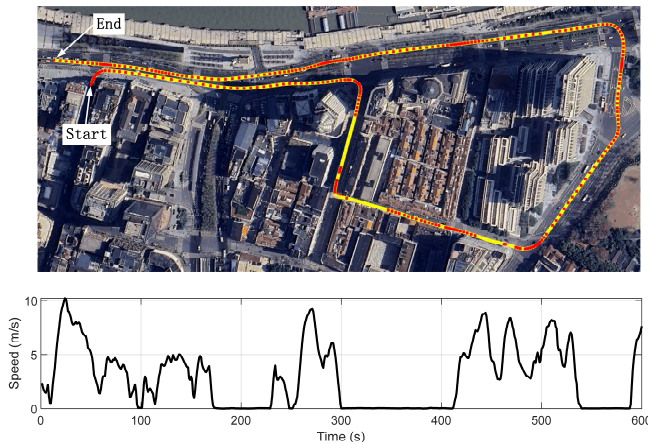

Fig. 12. The vehicle's trajectory (yellow) overlapped with the GNSS measurements (red dots) and the speed.

The initial attitude error [60, 180, 60] deg is added to the true attitude and the standard deviation is set to [180, 180, 180] deg. The attitude estimation errors are compared in Fig. 13, and the estimated biases are presented in Figs. 14-15, in which the largest biases are about 500 deg/h and 0.01g for the gyroscopes and accelerometers, respectively. The RMSEs of heading errors over specific intervals are compared in the left columns of the Table V. These results indicate that the performances of LQEKF-Iter, RQEKF-Iter and Clifford-RQEKF-Iter are similar, and they converge to similar accuracy with that of Ideal-EKF in about 15 s. As the results in Fig. 6, Fig. 16 also shows that the current methods cannot converge to the ground-truth lever arms using the velocity-only measurements, and the lever-arm estimation of RQEKF-Iter is less stable than the other methods. Therefore, the lever arms would better be pre-calibrated for low-cost INS/GNSS integrated navigation systems.

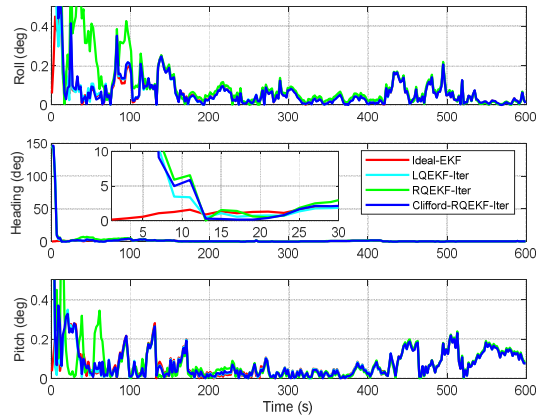

Fig. 13. The attitude estimation errors in the test.

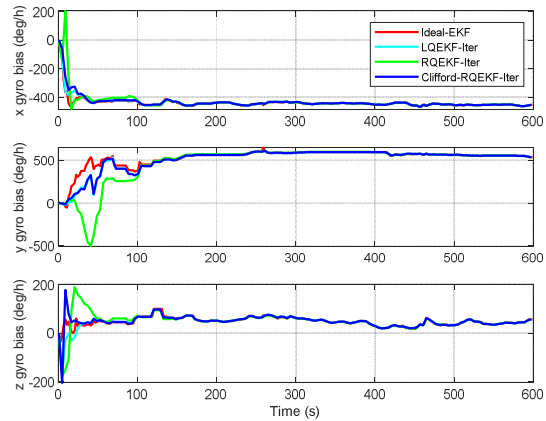

Fig. 14. The estimated gyroscope biases in the test.

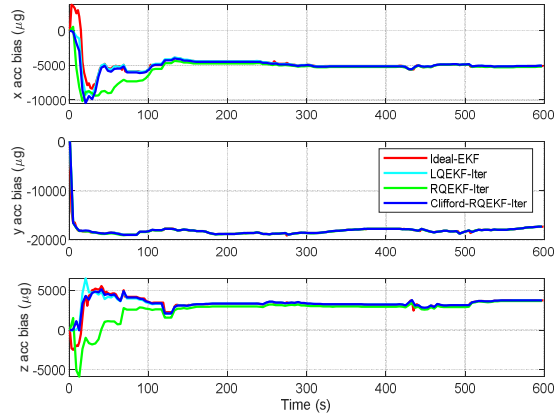

Fig. 15. The estimated accelerometer biases in the test.

**TABLE V**
**The RMSE of heading errors (deg) in the tests**

| Methods | 1~10 s | | 10~50 s | | 200~600 s | |
|---|---|---|---|---|---|---|
| Standard EKF | 0.88 | 3.62 | 1.56 | 2.50 | 0.76 | 0.75 |
| LQEKF-Iter | 95.5 | 125 | 1.88 | 59.7 | 0.74 | 70.1 |
| RQEKF-Iter | 93.3 | 113 | 4.67 | 66.5 | 0.78 | 14.4 |
| Clifford-RQEKF-Iter | 94.7 | 121 | 2.25 | 5.92 | 0.76 | 0.79 |

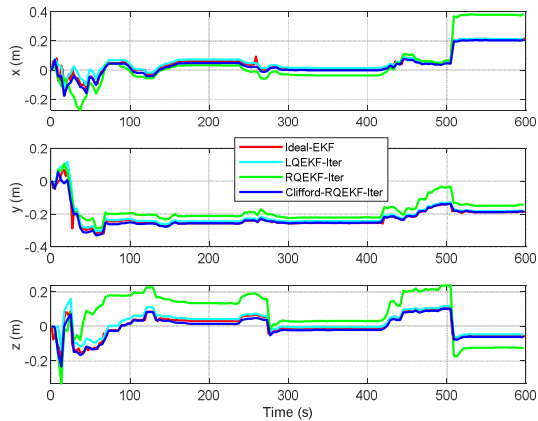

Fig. 16. The estimated lever arms in the test.

To test the filtering performance under larger IMU biases, the biases of the test data are artificially enlarged to 2000 deg/h and 0.1g for the gyroscopes and accelerometers, respectively. Results in Figs. 17-20 show that Clifford-RQEKF-Iter performs obviously better than LQEKF-Iter and RQEKF-Iter, and LQEKF-Iter and RQEKF-Iter fail to converge quickly comparing with the results under moderate biases in Figs. 13-16. The RMSEs of the heading errors over specific intervals are compared in the right columns of the Table V, which indicates that the heading errors in the transient and asymptotic phases of Clifford-RQEKF-Iter also become larger due to the increased biases, but the stabilized heading errors are similar. Extensive tests also reveal that LQEKF-Iter performs normally only if the biases can be estimated well from the beginning, which depends on the observability of biases and the quality of GNSS measurements. Otherwise, the covariance propagation (52) would contain significant errors caused by inaccurate IMU biases, making the accurate estimation of the attitude and biases harder. Comparing RQEKF-Iter with Clifford-RQEKF-Iter, it can be found that representing the biases and lever arms on the group $SE_{k+2}(3)$ improves the filtering performance in contrast with adopting the additive vector errors.

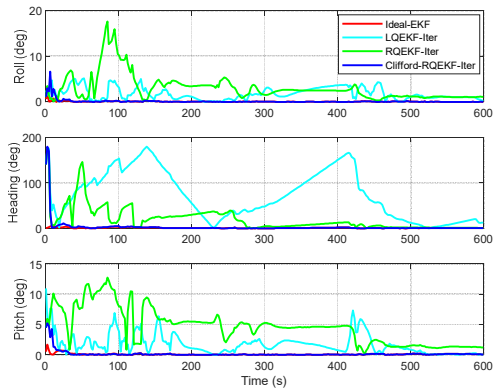

Fig. 17. The attitude estimation errors under intentionally enlarged biases in the test.

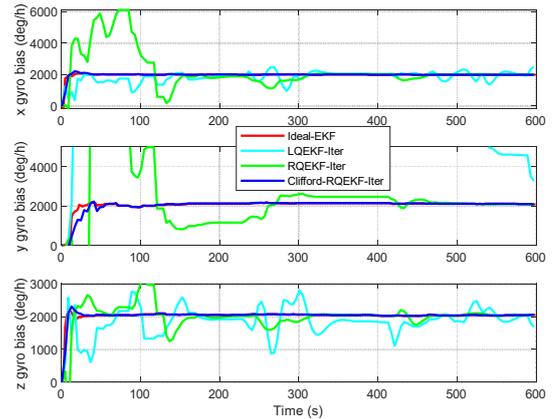

Fig. 18. The estimated gyroscope biases (intentionally enlarged).

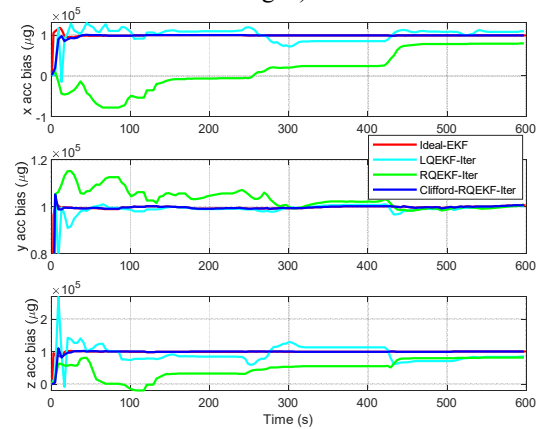

Fig. 19. The estimated accelerometer biases (intentionally enlarged).

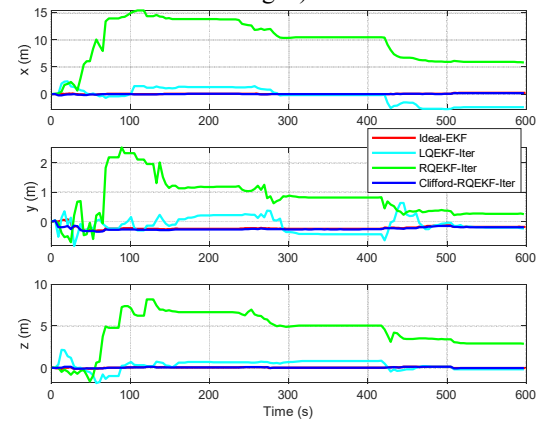

Fig. 20. The estimated lever arms under intentionally enlarged biases.

Note that the divergence of the traditional EKF is difficult to be detected online by checking the filtering consistency, since the estimation errors are inconsistent with the estimation covariance. It is also true for the diverged cases of LQEKF-Iter and RQEKF-Iter. Considering the large convergence domain of LQEKF-Iter and RQEKF-Iter, this work adopts the straightforward reset operation [62] to recover the diagonal proper covariance matrix of states. Specifically, the position and velocity are reinitialized by the current velocity and position provided by the GNSS measurements, and the others states are reset as zero vectors. The standard deviations of the attitude error are reset as [180, 180, 180] deg. Note that the current attitude is regarded as the initial attitude upon the reset, which is much closer to the truth in contrast with the assumed initial attitude error [60, 180, 60] deg. The results of applying the straightforward reset operation at 10 s for the LQRKF-Iter and RQEKF-Iter are compared in Figs. 21-23, respectively. It indicates that the direct reset can effectively obviate the filtering divergence. As shown in Fig. 24, however, the reset operation cannot avoid the divergence of EKF-Iter due to its limited convergence region. In contrast, the reset operation further expedites the convergence of Clifford-RQEKF-Iter at the beginning of the filtering process.

In most applications the constant IMU biases can be largely calibrated by keeping the IMU static in a short time. According to the results in Fig. 13, the LQEKF-Iter, RQEKF-Iter and Clifford-RQEKF-Iter can be directly applied to the state estimation of the low-cost INS/GNSS systems without requiring the knowledge of the prior attitude and the carrier's motion constraint, such as the nonholonomic constraint applied in [40], [41]. For the uncalibrated IMU, however, the Clifford-RQEKF-Iter is superior in dealing with large IMU biases.

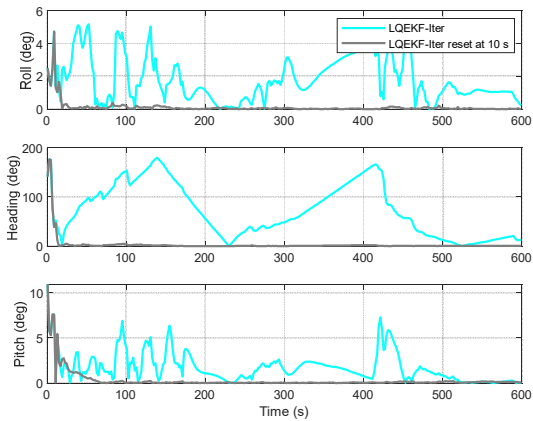

Fig. 21. The attitude estimation errors of LQEKF-Iter under intentionally enlarged biases with/without the direct reset at 10 s.

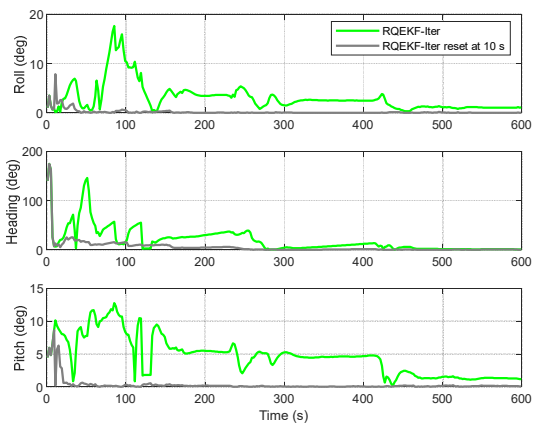

Fig. 22. The attitude estimation errors of RQEKF-Iter under intentionally enlarged biases with/without the direct reset at 10 s.

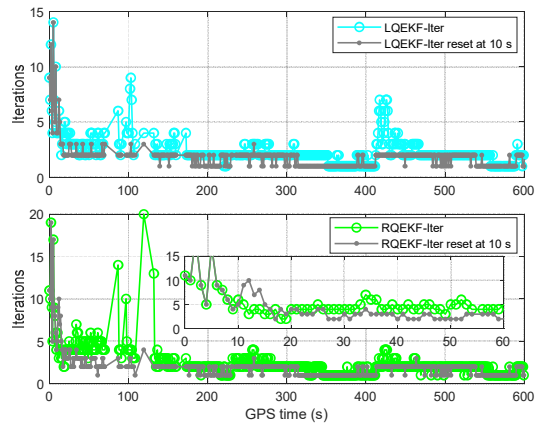

Fig. 23. The iteration numbers of LQEKF-Iter and RQEKF-Iter under intentionally enlarged biases with/without the direct reset at 10 s.

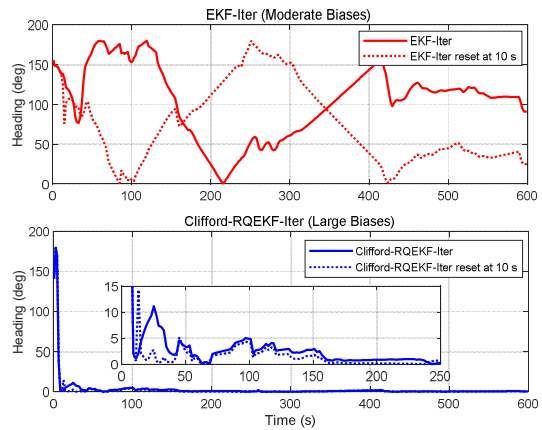

Fig. 24. Heading errors of EKF-Iter and Clifford-RQEKF-Iter under intentionally enlarged biases with/without the direct reset.

## VI. CONCLUSIONS

In the current work, the iterated extended Kalman filter based on the right-error Clifford algebra error is proposed for the low-cost INS/GNSS integrated navigation system. The Clifford algebra is introduced to represent the system states on the Special Euclidean group $SE_{k+2}(3)$, and its isomorphism with the matrix Lie group is verified. The quasi-group-affine system kinematics denoted by the Clifford algebra are applied to design the Clifford-RQEKF method, which is consistent with the rationale of the two-frames group invariant filtering theory. Finally, the iterated filtering method is exploited to eliminate the intractable linearization error in the measurement models, which is paramount for achieving the global convergence of the filters. Simulations and experiments for the low-cost INS/GNSS system show that the Clifford-RQEKF-Iter, LQEKF-Iter and RQEKF-Iter can converge without the initial attitude information, while the Clifford-RQEKF-Iter performs the best under extremely large IMU biases. This article corroborates that modeling the body-frame vectors on the Lie group is conducive to improve the state estimation accuracy especially when the consumer-grade IMU is poorly calibrated.

Further works will delve into integrating more optimization techniques to the Clifford algebra-based filtering method.

## APPENDIX A

This Appendix provides the derivations of the linearized error kinematic model in (41). The differential equation in (38) can be approximated as

$$\Delta \dot{\boldsymbol{\sigma}}_r = \Delta \boldsymbol{\omega}_r^e$$
$$= Ad_{\hat{q}_e^b}\left(\boldsymbol{\omega}_{ib}^b - \boldsymbol{C}_e^b \boldsymbol{\omega}_{ie}^e - \left(\tilde{\boldsymbol{\omega}}_{ib}^b - \hat{\boldsymbol{b}}_g - \hat{\boldsymbol{C}}_e^b \boldsymbol{\omega}_{ie}^e\right)\right)$$
$$= Ad_{\hat{q}_e^b}\left(\hat{\boldsymbol{b}}_g - \boldsymbol{b}_g + \left(\hat{\boldsymbol{C}}_e^b - \boldsymbol{C}_e^b \boldsymbol{C}_e^{\hat{e}}\right)\boldsymbol{\omega}_{ie}^e\right) - Ad_{\hat{q}_e^b}\boldsymbol{n}_g \quad (61)$$
$$= -\hat{\boldsymbol{C}}_b^e\left(\boldsymbol{b}_g - \hat{\boldsymbol{b}}_g\right) + \hat{\boldsymbol{C}}_b^e\left(\hat{\boldsymbol{C}}_e^b \Delta \boldsymbol{\sigma}_r \times \boldsymbol{\omega}_{ie}^e\right) - \hat{\boldsymbol{C}}_b^e \boldsymbol{n}_g$$
$$= -\boldsymbol{\omega}_{ie}^e \times \Delta \boldsymbol{\sigma}_r - \Delta \boldsymbol{b}_{g,r} - \hat{\boldsymbol{C}}_b^e \boldsymbol{n}_g$$

in which, $\tilde{\boldsymbol{\omega}}_{ib}^b = \boldsymbol{\omega}_{ib}^b + \boldsymbol{b}_g + \boldsymbol{n}_g, \hat{\boldsymbol{\omega}}_{ib}^b = \tilde{\boldsymbol{\omega}}_{ib}^b - \hat{\boldsymbol{b}}_g, \boldsymbol{C}_{\hat{e}}^e \approx 1 + \Delta \boldsymbol{\sigma}_r \times$.

The differential equation for the transformed velocity error in (38) is approximated as

$$\Delta \dot{\boldsymbol{v}}_r^e = Ad_{\Delta q_r^*}\left(Ad_{\hat{q}_e^b}\boldsymbol{f}^b - \boldsymbol{\omega}_{ie}^e \times \boldsymbol{v}^e + \boldsymbol{g}^e\right)$$
$$- \left(Ad_{\hat{q}_e^b}\hat{\boldsymbol{f}}^b - \boldsymbol{\omega}_{ie}^e \times \hat{\boldsymbol{v}}^e + \hat{\boldsymbol{g}}^e\right) + \left(Ad_{\Delta q_r^*}\boldsymbol{v}^e\right) \times \Delta \boldsymbol{\omega}_r^e$$
$$= Ad_{\hat{q}_e^b}\left(\boldsymbol{f}^b - \hat{\boldsymbol{f}}^b\right) - Ad_{\Delta q_r^*}\left(\boldsymbol{\omega}_{ie}^e \times \boldsymbol{v}^e\right)$$
$$+ \boldsymbol{\omega}_{ie}^e \times \left(Ad_{\Delta q_r^*}\boldsymbol{v}^e - \Delta \boldsymbol{v}_r^e\right) + Ad_{\Delta q_r^*}\boldsymbol{g}^e - \boldsymbol{g}^e$$
$$+ \left(Ad_{\Delta q_r^*}\boldsymbol{v}^e\right) \times \left(-\boldsymbol{\omega}_{ie}^e \times \Delta \boldsymbol{\sigma}_r - \Delta \boldsymbol{b}_{g,r} - \hat{\boldsymbol{C}}_b^e \boldsymbol{n}_g\right) \quad (62)$$
$$\approx \hat{\boldsymbol{C}}_b^e\left(\boldsymbol{f}^b - \hat{\boldsymbol{f}}^b\right) - \boldsymbol{\omega}_{ie}^e \times \Delta \boldsymbol{v}_r^e + \boldsymbol{g}^e \times \Delta \boldsymbol{\sigma}_r$$
$$- \boldsymbol{v}^e \times \Delta \boldsymbol{b}_{g,r} - \boldsymbol{v}^e \times \hat{\boldsymbol{C}}_b^e \boldsymbol{n}_g$$
$$= \boldsymbol{g}^e \times \Delta \boldsymbol{\sigma}_r - \boldsymbol{\omega}_{ie}^e \times \Delta \boldsymbol{v}_r^e - \boldsymbol{v}^e \times \Delta \boldsymbol{b}_{g,r} - \Delta \boldsymbol{b}_{a,r}$$
$$- \boldsymbol{v}^e \times \hat{\boldsymbol{C}}_b^e \boldsymbol{n}_g - \hat{\boldsymbol{C}}_b^e \boldsymbol{n}_a$$

in which, $\tilde{\boldsymbol{f}}^b = \boldsymbol{f}^b + \boldsymbol{b}_a + \boldsymbol{n}_a$, $\hat{\boldsymbol{f}}^b = \tilde{\boldsymbol{f}}^b - \hat{\boldsymbol{b}}_a$, and the geometric product of two vector quaternions satisfies $(\boldsymbol{ab} - \boldsymbol{ba})/2 = \boldsymbol{a} \times \boldsymbol{b}$. Note that the gravity error caused by the position estimation error is neglected in contrast with the derivations in [42].

The differential equation for the position error is computed by

$$\Delta \dot{\boldsymbol{r}}_r^e = Ad_{\Delta q_r^*}\left(\boldsymbol{v}^e - \boldsymbol{\omega}_{ie}^e \times \boldsymbol{r}^e\right) - \hat{\boldsymbol{v}}^e + \boldsymbol{\omega}_{ie}^e \times \hat{\boldsymbol{r}}^e$$
$$+ Ad_{\Delta q_r^*}\boldsymbol{r}^e \times \Delta \boldsymbol{\omega}_r^e$$
$$= Ad_{\Delta q_r^*}\boldsymbol{v}^e - \hat{\boldsymbol{v}}^e - Ad_{\Delta q_r^*}\left(\boldsymbol{\omega}_{ie}^e \times \boldsymbol{r}^e\right)$$
$$+ Ad_{\Delta q_r^*}\boldsymbol{r}^e \times \left(-\boldsymbol{\omega}_{ie}^e \times \Delta \boldsymbol{\sigma}_r - \Delta \boldsymbol{b}_{g,r} - \hat{\boldsymbol{C}}_b^e \boldsymbol{n}_g\right) \quad (63)$$
$$+ \boldsymbol{\omega}_{ie}^e \times \left(Ad_{\Delta q_r^*}\boldsymbol{r}^e - \Delta \boldsymbol{r}_r^e\right)$$
$$\approx \Delta \boldsymbol{v}_r^e - \boldsymbol{\omega}_{ie}^e \times \Delta \boldsymbol{r}_r^e - \boldsymbol{r}^e \times \Delta \boldsymbol{b}_{g,r} - \boldsymbol{r}^e \times \hat{\boldsymbol{C}}_b^e \boldsymbol{n}_g$$

## APPENDIX B

Similar to (21), the iterated error states $\Delta \boldsymbol{x}_r = \delta \boldsymbol{\eta}_{i+1}$ encodes the Lie algebra of $\Delta \breve{q}_r$, i.e., $\Delta \breve{q}_r = \exp(\Delta \boldsymbol{x}_r)$ and $\Delta q_r = \exp(\Delta \boldsymbol{\sigma}_r)$. The state errors in (29) can be alternatively represented by the following matrix Lie group, which is isomorphic to $\Delta \breve{q}_r$.

$$\hat{\chi}\chi^{-1} = \begin{bmatrix} \boldsymbol{C}_e^{\hat{e}} & \Delta \boldsymbol{v}_r^e & \Delta \boldsymbol{r}_r^e & \Delta \boldsymbol{b}_{g,r} & \Delta \boldsymbol{b}_{a,r} & \Delta \boldsymbol{l}_r^b \\ \boldsymbol{0}_{5\times 3} & & & \boldsymbol{I}_5 & & \end{bmatrix}_{8\times 8} \quad (64)$$

And,

$$\chi = \begin{bmatrix} \boldsymbol{C}_b^e & \boldsymbol{v}^e & \boldsymbol{r}^e & \boldsymbol{C}_b^e \boldsymbol{b}_g & \boldsymbol{C}_b^e \boldsymbol{b}_a & \boldsymbol{C}_b^e \boldsymbol{l}^b \\ \boldsymbol{0}_{5\times 3} & & & \boldsymbol{I}_5 & & \end{bmatrix}_{8\times 8} \quad (65)$$

which is isomorphic to the Clifford algebra in (25).

According to [35], [54], the state errors in (64) can be computed by the estimated error states $\Delta \boldsymbol{x}_r$ in (40) by

$$\boldsymbol{C}_e^{\hat{e}} = \exp(-\Delta \boldsymbol{\sigma}_r), \quad \Delta \boldsymbol{v}_r^e = \boldsymbol{J}_l(-\Delta \boldsymbol{\sigma}_r)\Delta \boldsymbol{\sigma}_v,$$
$$\Delta \boldsymbol{r}_r^e = \boldsymbol{J}_l(-\Delta \boldsymbol{\sigma}_r)\Delta \boldsymbol{\sigma}_p, \quad \Delta \boldsymbol{b}_{g,r} = \boldsymbol{J}_l(-\Delta \boldsymbol{\sigma}_r)\Delta \boldsymbol{\sigma}_{b_g}, \quad (66)$$
$$\Delta \boldsymbol{b}_{a,r} = \boldsymbol{J}_l(-\Delta \boldsymbol{\sigma}_r)\Delta \boldsymbol{\sigma}_{b_a}, \quad \Delta \boldsymbol{l}_r^b = \boldsymbol{J}_l(-\Delta \boldsymbol{\sigma}_r)\Delta \boldsymbol{\sigma}_l$$

in which, the left-Jacobian matrix of the rotation vector is [54]

$$\boldsymbol{J}_l(\boldsymbol{\theta}) = \frac{\sin\theta}{\theta}\boldsymbol{I}_3 + \left(1 - \frac{\sin\theta}{\theta}\right)\boldsymbol{nn}^T - \frac{1-\cos\theta}{\theta}\boldsymbol{n}\times \quad (67)$$

and, it satisfies $\exp(\boldsymbol{\theta}\times)\boldsymbol{J}_l(-\boldsymbol{\theta}) = \boldsymbol{J}_l(\boldsymbol{\theta})$.

Substituting (66) into (29), the states are updated by

$$q_e^b = \exp(\Delta \boldsymbol{\sigma}_r/2)\hat{q}_e^b,$$
$$\boldsymbol{v}^e = \boldsymbol{C}_{\hat{e}}^e \hat{\boldsymbol{v}}_r^e + \boldsymbol{J}_l(\Delta \boldsymbol{\sigma}_r)\Delta \boldsymbol{\sigma}_v,$$
$$\boldsymbol{r}^e = \boldsymbol{C}_{\hat{e}}^e \hat{\boldsymbol{r}}^e + \boldsymbol{J}_l(\Delta \boldsymbol{\sigma}_r)\Delta \boldsymbol{\sigma}_p,$$
$$\boldsymbol{b}_g = \hat{\boldsymbol{b}}_g + \hat{\boldsymbol{C}}_b^{eT}\boldsymbol{J}_l(-\Delta \boldsymbol{\sigma}_r)\Delta \boldsymbol{\sigma}_{b_g}, \quad (68)$$
$$\boldsymbol{b}_a = \hat{\boldsymbol{b}}_a + \hat{\boldsymbol{C}}_b^{eT}\boldsymbol{J}_l(-\Delta \boldsymbol{\sigma}_r)\Delta \boldsymbol{\sigma}_{b_a},$$
$$\boldsymbol{l}^b = \hat{\boldsymbol{l}}^b + \hat{\boldsymbol{C}}_b^{eT}\boldsymbol{J}_l(-\Delta \boldsymbol{\sigma}_r)\Delta \boldsymbol{\sigma}_l$$

in which, the following relationship is used

$$\boldsymbol{C}_{\hat{e}}^e \boldsymbol{J}_l(-\Delta \boldsymbol{\sigma}_r) = \exp(\Delta \boldsymbol{\sigma}_r \times)\boldsymbol{J}_l(-\Delta \boldsymbol{\sigma}_r) \quad (69)$$
$$= \boldsymbol{J}_l(\Delta \boldsymbol{\sigma}_r)$$